\documentclass{aa} 
 
\usepackage{graphicx}
\usepackage{txfonts}
\usepackage{natbib}
\bibpunct{(}{)}{;}{a}{}{,}

\begin{document}

\title{Effect of the initial mass function on the dynamical SMBH mass estimate in the nucleated early-type galaxy FCC\,47}

   \author{Sabine Thater\inst{1} \and Mariya Lyubenova\inst{2} \and Katja Fahrion\inst{3} \and Ignacio Mart{\'i}n-Navarro\inst{4,5} \and Prashin Jethwa\inst{1} \and \\ Dieu D. Nguyen\inst{6} \and Glenn van de Ven\inst{1}}
   
   \institute{Department of Astrophysics, University of Vienna, T\"urkenschanzstraße 17, 1180 Vienna, Austria\\
              \email{sabine.thater@univie.ac.at}
         \and
            European Southern Observatory, Karl-Schwarzschild-Straße 2, 85748 Garching bei München, Germany
        \and  
                European Space Agency, European Space Research and Technology Centre, Keplerlaan 1, 2200 AG Noordwijk, Netherlands
            \and    
                Instituto de Astrof{\'i}sica de Canarias, Calle Via L{\'a}ctea s/n, 38200 La Laguna, Tenerife, Spain
           \and
            Depto. Astrof{\'i}sica, Universidad de La Laguna, Calle Astrof{\'i}sico Francisco S{\'a}nchez s/n, 38206 La Laguna, Tenerife, Spain
        \and
            Universit\'e de Lyon 1, Ens de Lyon, CNRS, Centre de Recherche Astrophysique de Lyon (CRAL) UMR5574, F-69230 Saint-Genis-Laval, France\\
}

   \date{Received: 2 November 2022/ Accepted: 14 April 2023}

 \abstract
{Supermassive black holes (SMBHs) and nuclear star clusters (NSCs) co-exist in many galaxies. While the formation history of the  black hole is essentially lost, NSCs preserve their evolutionary history imprinted onto their stellar populations and kinematics. Studying SMBHs and NSCs in tandem might help us to ultimately reveal the build-up of galaxy centres. In this study, we combine large-scale VLT/MUSE and high-resolution adaptive-optics-assisted VLT/SINFONI observations of the early-type galaxy FCC\,47 with the goal being to assess the effect of a spatially (non-)variable initial mass function (IMF) on the determination of the mass of the putative SMBH in this galaxy. We achieve this by performing DYNAMITE Schwarzschild orbit-superposition modelling of the galaxy and its NSC. In order to properly take account of the stellar mass contribution to the galaxy potential, we create mass maps using a varying stellar mass-to-light ratio derived from single stellar population models with fixed and with spatially varying IMFs. Using the two mass maps, we estimate black hole masses of $(7.1^{+0.8}_{-1.1})\times 10^7\,M_{\odot}$ and $(4.4^{+1.2}_{-2.1}) \times 10^7\,M_{\odot}$ at $3\sigma$ signifance, respectively. Compared to models with constant stellar-mass-to-light ratio, the black hole masses decrease by 15\% and 48\%, respectively. Therefore, a varying IMF, both in its functional form and spatially across the galaxy, has a non-negligible effect on the SMBH mass estimate. Furthermore, we find that the SMBH in FCC\,47 has probably not grown over-massive compared to its very over-massive NSC. }


\titlerunning{The centre of the nucleated galaxy FCC\,47}
   
\authorrunning{Thater et al.}

\maketitle


\section{Introduction}\label{s:intro}

Nuclear star clusters (NSCs) are amongst the densest stellar systems in the Universe. They are typically found in the centres of galaxies with stellar masses of between 10$^8$\,M$_{\odot}$ and 10$^{10}$\,M$_{\odot}$ (see review by \citealt{Neumayer2020} as well as the NSC compilation by \citealt{Hoyer2021}). More massive galaxies with stellar masses of at least $10^{11}$\,M$_{\odot}$ lack NSCs and only harbour supermassive black holes (SMBHs) in their centres (see reviews by \citealt{Kormendy2013} and \citealt{Antonini2015}).
In the centres of some galaxies, massive black holes were found to live in co-existence with their NSCs \citep[e.g.][]{Genzel2010,Filippenko2003,Seth2008,Graham2009,Neumayer2012}. Therefore, dynamical interactions between NSCs and SMBHs are expected. Studying NSCs and SMBH in tandem might help us to reveal the build-up of galaxy centres, as NSCs preserve their own evolutionary history as imprinted onto their stellar populations and kinematics. Consequently, understanding the evolution of NSCs might ultimately provide important indications as to the pathways that SMBHs haven taken to grow. Several studies investigated the mass ratio between SMBHs and NSCs and noticed an increase with host galaxy mass \citep{Graham2009, Seth2008, Neumayer2012, Georgiev2016, Nguyen2018, Neumayer2020}. However, the small number of robust measurements and scatter spanning multiple magnitudes in this ratio have made it difficult to derive conclusions about the evolution of galaxy centres. \\

In this publication, we add a new galaxy to the SMBH-NSC compilations. FCC\,47 (NGC 1336) is an early-type galaxy (ETG) located in the Fornax cluster at $\sim$20 Mpc. We list the basic properties of FCC\,47 in Table \ref{properties}.
FCC\,47 was included in the \textit{Hubble} Space Telescope (HST) Advanced Camera for Surveys (ACS) Fornax Cluster Survey \citep[ACSFCS;][]{Jordan2007} and its NSC was evident from photometric decomposition by \cite{Turner2012}. \cite{Turner2012} derived the effective radius of the NSC to be $r_{\rm eff,NSC} = 0.75 \pm 0.13 \arcsec$ in
the F475W filter, which corresponds to 66.5 pc at a distance of 18.3 Mpc. Using integral-field spectroscopy data from the Multi Unit Spectroscopic Explorer (MUSE), \cite{Fahrion2019} found that the NSC has a mass of about $7 \times 10^{8}$ M$_{\odot}$. Compared to the mass of the host galaxy ($M_{\rm *,gal}\sim10^{10}$ M$_{\odot}$), this NSC is very over-massive. The NSC
also appears very pronounced in kinematic data due to its strong rotation \citep{Lyubenova2019, Fahrion2019} and a comparison to the rotation of the main galaxy body reveals that the NSC is a kinematically decoupled component with a rotation axis almost orthogonal to the short axis of the host galaxy. From its star formation history \citep{Fahrion2019}, we know that the NSC in FCC\,47 is very old ( > 13 Gyr) compared to the host galaxy (about 8 Gyr). The NSC therefore likely formed a long time ago, with no significant evolution within the NSC following its initial formation. The NSC in FCC\,47 provides a well-studied test case to explore the evolution of galaxy centres. 

\cite{Lee2019} searched for nuclear X-ray signal in FCC 47 but could not identify any nuclear activity ($\log L_{\rm x} < 38.48$ erg s$^{-1}$). Based on optical spectroscopy, \cite{Zaw2019} also classified FCC 47 as a galaxy with a non-active nucleus. On the other hand, FCC 47 was identified as a radio source ($S_{1.4}=2.5$ mJy) in the NRAO VLA Sky Survey \citep{Condon2019} and was classified as an active galaxy nucleus (AGN) based on radio and infrared data. 

The goal of the present study is to test whether or not we can dynamically measure a black hole in this NSC and whether or not this black hole is as over-massive as the NSC. Based on the mass of FCC\,47 ($M_{\rm *,gal}\sim10^{10}$ M$_{\odot}$), the $M_{\rm BH} - M_{\rm *,gal}$ relation for ETGs by \cite{Lelli2022} predicts a central black hole of $M_{\rm BH} \sim 10^7 M_{\odot}$. The sphere of influence\footnote{The radius of the sphere of influence is defined as $R_{\rm SoI}=GM_{\rm BH} / \sigma^2_{\rm e,star}$ where G is the gravitational constant. Within $R_{\rm SoI}$ the gravitational potential is dominated by the SMBH.} (SoI) of such a black hole is about 5 pc, which is about 0.04 arcsec at a distance of $18.3 \pm 0.6$ Mpc \citep{Blakeslee2009}. \citet{Thater2017} showed that such a spatial resolution allows a robust estimate of the central black hole mass if high-spatial-resolution integral field spectroscopic observations are modelled with the Schwarzschild orbit-superposition technique.

In the past, due to the limited data coverage, most dynamical models of NSCs and their massive black holes were performed with Jeans modelling (e.g. \citealt{Walcher2005,Neumayer2012, Feldmeier2014, Nguyen2017a, Nguyen2018,Nguyen2019a}) or tilted ring modelling (e.g. \citealt{denBrok2015}). \cite{Lyubenova2013} bridged this limitation by combining adaptive-optics-assisted Spectrograph for INtegral Field Observations in the Near Infrared (SINFONI) kinematics with long-slit kinematics and constructed triaxial \cite{Schwarzschild1979} orbit-superposition models for FCC 277. Since then, only a handful of NSCs have been modelled with Schwarzschild modelling. \cite{FeldmeierKrause2017} modelled the NSC in the centre of the Milky Way (MW) and recovered the mass of the central massive black hole, Sgr A*. Subsequently, \cite{Fahrion2019} used MUSE observations to model the complicated dynamical structure of FCC\,47. Even though the NSC is formally resolved in the MUSE data, the spatial resolution of 0.7\arcsec  allows neither a detailed investigation of its orbital structure nor the mass measurement of the putative SMBH. For this reason, we extend the work by \cite{Fahrion2019} in this study and combine the large-scale MUSE observations with high-resolution adaptive-optics-assisted SINFONI observations with the aim being  to probe the very centre of FCC\,47 and estimate the mass of its  black hole. 
 
 Dynamical modelling of NSCs needs to be performed very carefully. NSCs often have a very different stellar mass-to-light ratio ($M_*/L$) from the host galaxy (e.g. \citealt{Georgiev2016}). Therefore, in some dynamical studies, the NSC was added as an additional spherical component with a separate $M_*/L$ (e.g. \citealt{Nguyen2017a,Nguyen2019a}). The $M_*/L$ gradient in those studies resulted from variations in age and metallicity but a fixed IMF (\citealt{Salpeter1955}, \citealt{Kroupa2001} or \citealt{Chabrier2003}). In this study, we extend previous work by creating dynamical Schwarzschild models assuming a varying $M_*/L$ due to (1) differences in the stellar populations due to age and metallicity only and (2) differences in the stellar populations that include spatial IMF variations. \textit{This study provides the first SMBH mass estimation that takes into account a varying  $M_*/L$ due to spatial variations in the IMF.} 

The paper is organised as follows. In Section \ref{s:data}, we present the adaptive-optics-assisted SINFONI and MUSE integral-field spectroscopic observations, and the HST imaging data. In Section \ref{s:mge}, we derive the stellar mass distribution of the galaxy in three different ways: using (1) a constant stellar $M_*/L$ as commonly assumed, (2) a spatially varying $M_*/L$ and fixed MW-like IMF, or (3) a spatially varying $M_*/L$ and varying IMF. In Section \ref{s:dyna}, using the DYNAMITE software \citep{Jethwa2020,Thater2022b}, we perform triaxial Schwarzschild dynamical modelling of the stellar kinematics to constrain the intrinsic shape, dark matter, and SMBH mass of FCC 47. In Section \ref{s:discussion}, we discuss the effects of a (non)-variable IMF and put our results into context with the SMBH and NSC scaling relations. We conclude and summarise this paper in Section 6.

\section{Data}
\label{s:data}

\begin{table}
\caption{Basic properties of FCC~\,47 and its nuclear star cluster.}
\centering
\begin{tabular}{lcc}
\hline\hline
FCC\,47 &   &  Notes \\
\hline
Morphological type     &  S0   & 1   \\
Distance [Mpc]   &  $18.3 \pm 0.6 $ &  2    \\
Physical scale [pc arcsec$^{-1}$] & $88.7$ \\
Inclination [ $^{\circ}$] & $47 \pm 4$ & 3 \\
Effective radius [kpc]    & 2.60  & 4    \\
$\sigma_{\rm e,star}$ [km s$^{-1}$]    & $95 \pm 5$ & 5  \\
$\sigma_{\rm 0,star}$ [km s$^{-1}$]    &  $106 \pm 5$ & 5  \\ 
Galaxy stellar mass [M$_{\odot}$]   &  (1.0 $\pm 0.1) \times 10^{10} $ & 6  \\
\hline\hline
NSC &   &  \\
\hline
Effective radius [pc]    &  $66.5 \pm 11.1$  & 7    \\
NSC mass [M$_{\odot}$]   &  (7$\pm 1.4) \times 10^{8} $ & 8  \\
\hline
\end{tabular}
\\
\tablefoot
{1:  \cite{deVaucouleurs1991rc3}. 2: \cite{Blakeslee2009} using surface-brightness fluctuation measurements. 3: The inclination corresponds to the $\theta$ viewing angle of our triaxial Schwarzschild orbit-superposition models. 4: Derived by applying the routine mge{\_}half{\_}light{\_}isophote of the Jeans Anisotropic Modelling package \citep{Cappellari2008} on our surface brightness model (Section 3.1). For comparison, \cite{Ferguson1989} found 2.66 kpc.  5: Effective velocity dispersion and central velocity dispersion derived by co-adding the spectra of the large-scale optical IFU data in elliptical annuli of the size of the effective radius and 1 arcsec, respectively. 6: The galaxy stellar mass ($M_{\rm *,gal}$) was derived using the dynamical Schwarzschild orbit-superposition model with varying stellar populations and varying IMF. The mass does not change significantly for the other mass models. For comparison, \cite{Liu2019} derived $M_{\rm *,gal}=9.3 \times 10^9$ M$_{\odot}$ using a colour-based stellar $M_*/L$ and \cite{Fahrion2019} derived $M_{\rm *,gal}=1.6 \times 10^{10}$ M$_{\odot}$ using triaxial Schwarzschild models. 7: \cite{Turner2012} fitted a double Sersi\'c profile fit to the HST F475W filter image. 8: \cite{Fahrion2019} using stellar population models.}
\label{properties}
\end{table}

\begin{figure*}
  \centering
      \includegraphics[width=0.89\textwidth]{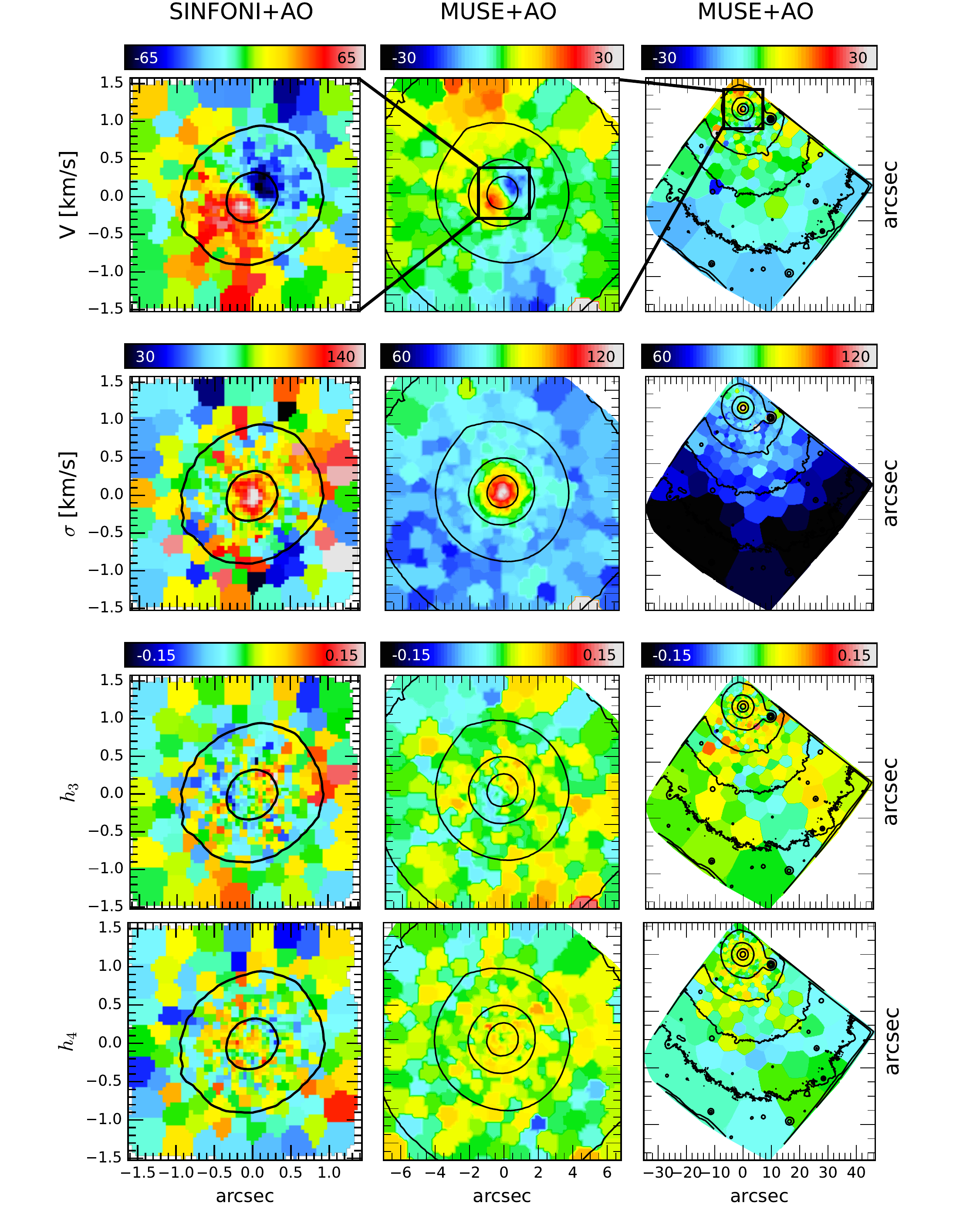} 

      \caption{Stellar kinematic maps of the galaxy FCC\,47 and its NSC overplotted with the isophotes of the IFU white-light images. The panels show from top to bottom the mean velocity, velocity dispersion, and $h_3$ and $h_4 $ Gauss-Hermite higher-order velocity moments. From left to right, we show the central $3\times 3$ arcsec extracted from the SINFONI observations, and the MUSE kinematics within the central $14''\times 14''$ and across the $60''\times 60''$ MUSE FOVs. SINFONI and MUSE kinematics are shown with different mean velocity and velocity dispersion scales due to the different spatial resolution. The host galaxy rotates significantly slower than the NSC ($r_{\rm eff}=0.7''$) in the centre of FCC\,47. North is up and east to the left. 
      }
      \label{ff:kin}
\end{figure*}

\subsection{SINFONI spectroscopic data}
We used the adaptive-optics(AO)-assisted SINFONI observations previously presented in \cite{Lyubenova2019}, without any modifications. The observations were performed under the science programme 092.B-0892 (PI: Lyubenova) in November 2013. In order to obtain the best possible resolution, the observations were carried out using laser-guide-star AO. The SINFONI data has a field of view (FOV) of $3'' \times 3''$ at a spatial sampling of $0.05''\times 0.10''$  and a spatial resolution of $0\farcs14$ at full width at half maximum (FWHM; see Appendix A for details). The data cube was spatially Voronoi-binned \citep{Cappellari2003} to a target signal-to-noise ratio (S/N) of 40 
\AA$^{-1}$  resulting in 404 
bins. \cite{Lyubenova2019} ran the penalised PiXel-Fitting \citep[pPXF;][]{Cappellari2004} on the binned spectra to derive the line-of-side velocity distribution (LOSVD) parametrised in rotational velocity, velocity dispersion, and $h_3$ and $h_4$ Gauss-Hermite moments that describe the skewness and excess kurtosis of the distribution. These authors fitted the wavelength range between 2.1 and 2.36 $\mu$m using seven template spectra of K and M giant stars but masking the regions affected by strong near-infrared sky lines. The resulting kinematic maps are presented in Fig. \ref{ff:kin}.

\subsection{MUSE spectroscopic data}
In order to map the large-scale kinematics, we used AO-assisted wide-field mode Multi-Unit Spectroscopic Explorer \citep[MUSE;][]{Bacon2010} kinematics of FCC\,47 that were previously presented in \cite{Fahrion2019}. The observations were obtained under the science programme 60.A-9192 (PI: Fahrion) at the Very Large Telescope during the science verification phase of the AO-assisted wide-field mode in September 2017. The IFU data have a FOV of $60'' \times 60''$ at a spatial sampling of $0.2''$ and a spatial resolution of $0.7''$ (FWHM). 

 As we wish to include stellar population properties in our model, we rebinned and re-extracted the stellar kinematics of the MUSE dataset. Using the Voronoi-binning scheme, the MUSE data were binned to a target S/N of 100 \AA$^{-1}$ resulting in 452 bins. This target S/N was chosen to ensure the accurate extraction of the LOSVD as well as stellar age, metallicity, $\alpha$ abundance, and IMF slope. The data were then fitted with the pPXF method to extract the parameters of the LOSVD ($V$, $\sigma$, $h_3$ and $h_4$) as well as stellar population properties such as age and metallicity. Compared to \cite{Fahrion2019}, no significant differences were found in our new stellar kinematics. The MUSE kinematic maps are also shown in Fig. \ref{ff:kin}.

\subsection{Photometric data}
We obtained HST Advanced Camera for Survey F850LP band imaging data (PI: Jord\'an; ACSFCS project; PID: 10217) for FCC\,47 from the ESA Hubble Science archive. ACS imaging data have the great advantage of a large covering area of $202'' \times  202''$, a spatial sampling of $0.05''$, and high spatial resolution ($\approx 0.09\arcsec$ FWHM). The large covering area is needed for creating an accurate orbit library of FCC\,47 for the dynamical Schwarzschild models, while the high spatial resolution is crucial for recovering the NSC and resolving the kinematics that are influenced by the central black hole. The downloaded F850LP image was calibrated with standard HST calibration routines. It has an exposure time of 1220 seconds such that we can also recover the lower surface brightness edges of the galaxy. The sensitivity limit of the F850LP HST image is discussed in detail in \cite{Jordan2007}. A two-dimensional comparison between the FoV of the F475W ACS image and the MUSE observations is shown in Figure 1 of \citet{Fahrion2019}. Our F850LP ACS image has the same coverage as the F475W ACS image.

\begin{figure}
  \centering
      \includegraphics[width=0.49\textwidth]{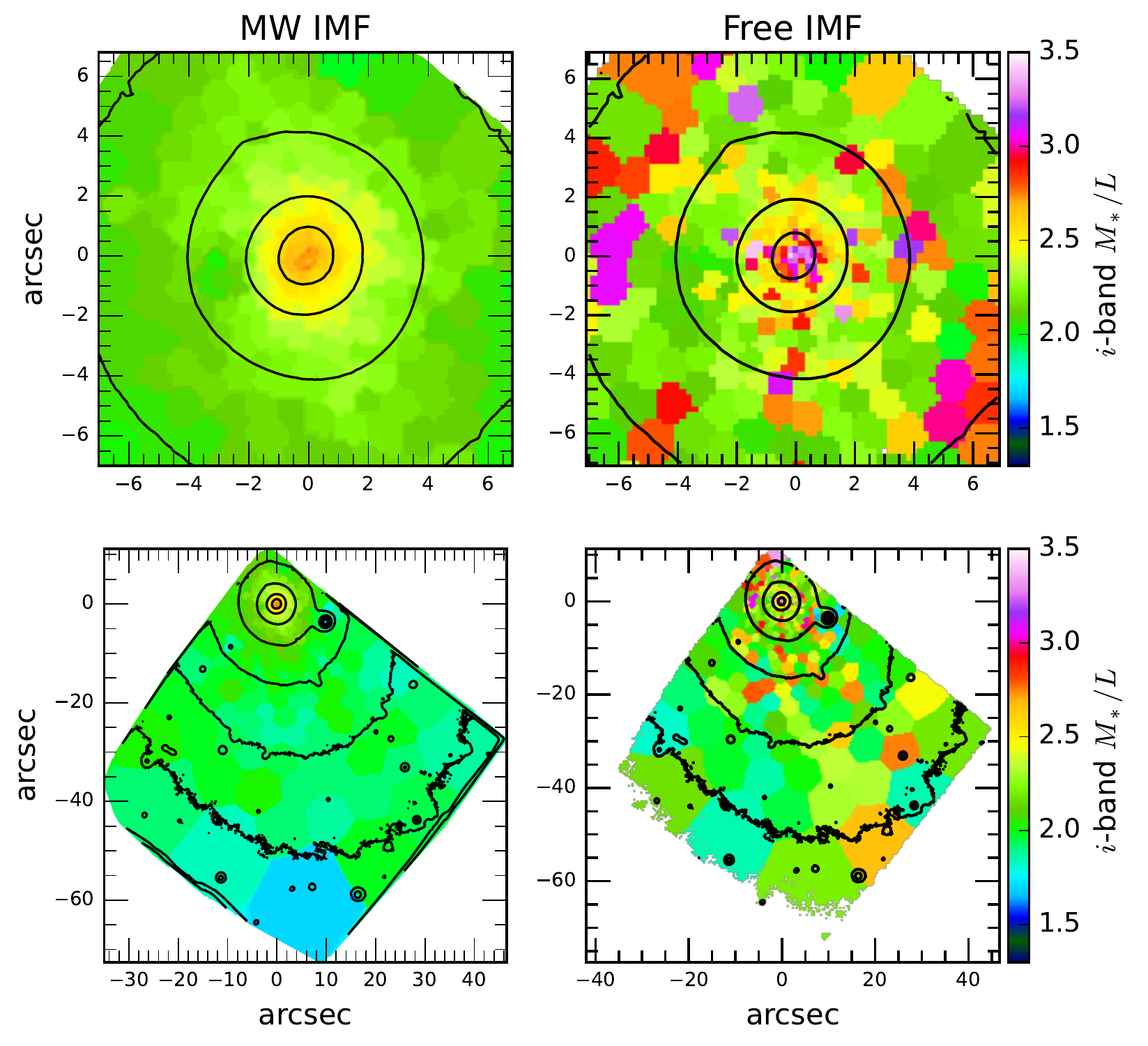}

      \caption{Stellar mass-to-light ratio ($M_*/L$) maps in the i-band derived from stellar population models fit to the MUSE observations. The left and right panels show the effect of different IMF assumptions. Left:  $M_{\rm *,MW-IMF}/L$ map derived from stellar population synthesis assuming a MW-like IMF (see Section 3.2). Right:  $M_{\rm *,free-IMF}/L$ map derived using a varying IMF but not allowing for IMF slopes lower than 0.8 (see Section 3.3). Top and bottom:  $M_*/L_{\rm i}$ map extracted from the MUSE observations shown within a zoom onto the centre of and across the full MUSE FoV, respectively. The $M_*/L$ is significantly higher in the centre of FCC 47 ---where the NSC is located--- than in the rest of the galaxy. For comparison, the dynamical model with constant mass-to-light ratio $M_{\rm *,const}/L$ is best fitted with 1.82 M$_{\odot}$/L$_{\odot}$ (see Section 4). Radial profiles of these maps are shown in Fig. \ref{ff:ml_gradient}.}
      
      \label{ff:ml}
\end{figure}

\section{Stellar mass models}
\label{s:mge}
The dynamical modelling of stellar kinematics requires a detailed description of the gravitational potential of the galaxy. We inferred the stellar mass distribution directly from the luminosity of the galaxy multiplied by its (spatially varying) mass-to-light ratio ($M_*/L$). 
FCC\,47 harbours a large NSC in its centre that has a very different $M_*/L$ from the host galaxy due to variations in the stellar populations and IMF (see Figure \ref{ff:ml}). A constant $M_*/L$ as commonly assumed in dynamical modelling is therefore an inadequate assumption for FCC\,47. In the following sections, we present three different mass models that we derived for FCC\,47: (1) using a traditional constant mass-to-light ratio $M_{\rm *,const}/L$; (2) using a varying mass-to-light ratio with fixed MW-like IMF ($M_{\rm *,MW-IMF}/L$); and (3) using a varying mass-to-light ratio and varying IMF ($M_{\rm *,free-IMF}/L$). From these three mass models we constructed three different sets of dynamical Schwarzschild models to describe the NSC and derive the SMBH mass. This is the first paper that accounts for IMF-variation-induced changes in $M_*/L$  when creating dynamical galaxy models. All $M_*/L$ discussed in the following sections are in the i-band. Ideally, we would like to use the z-band directly, but the wavelength range of MUSE does not cover the full F850LP filter. For this reason, we chose the i-band as it is reasonably close to the F850LP filter.

\begin{figure*}
  \centering
      \includegraphics[width=1\textwidth]{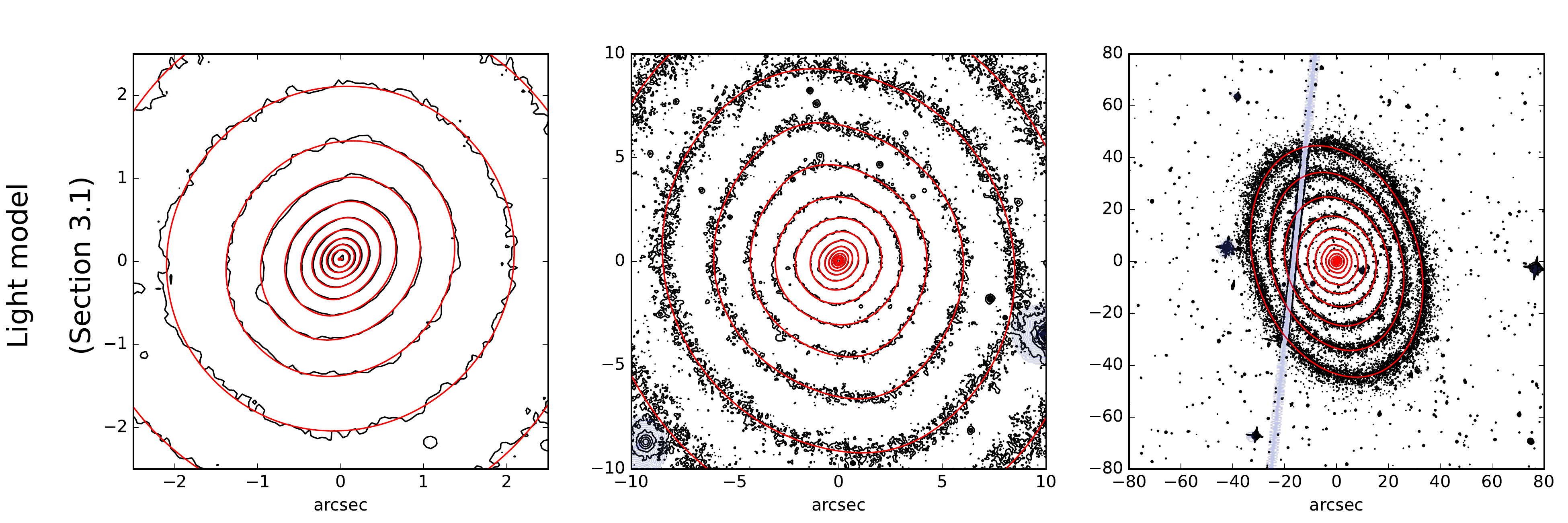}
     \includegraphics[width=1\textwidth]{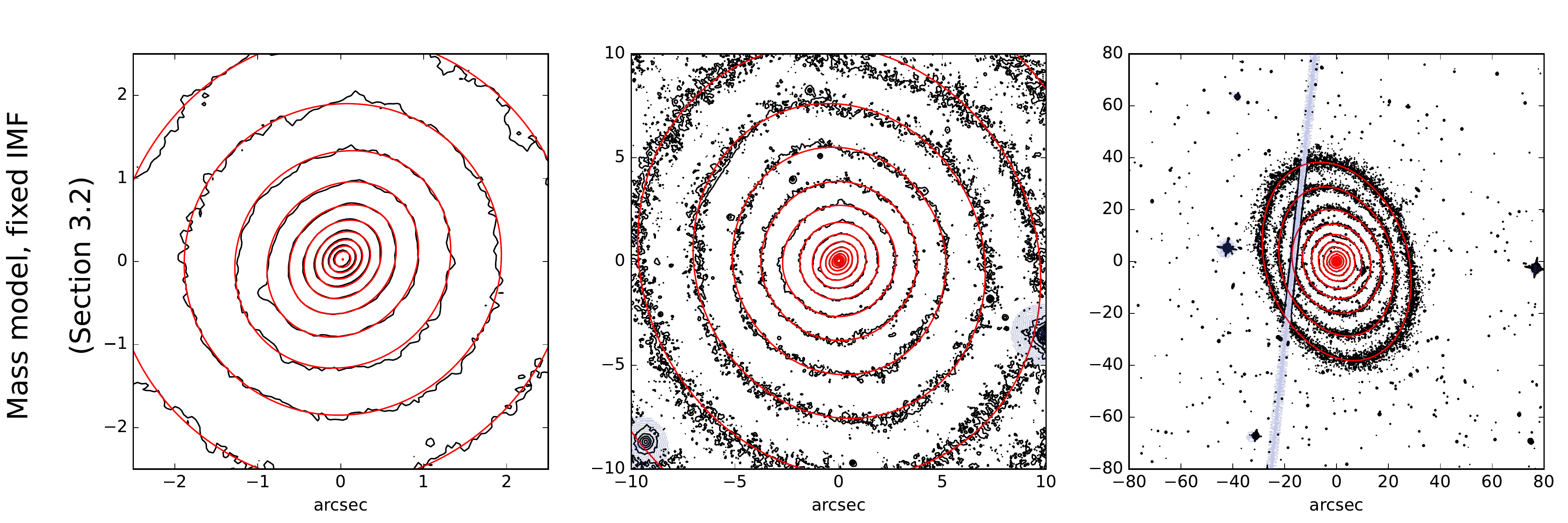}
      \includegraphics[width=1\textwidth]{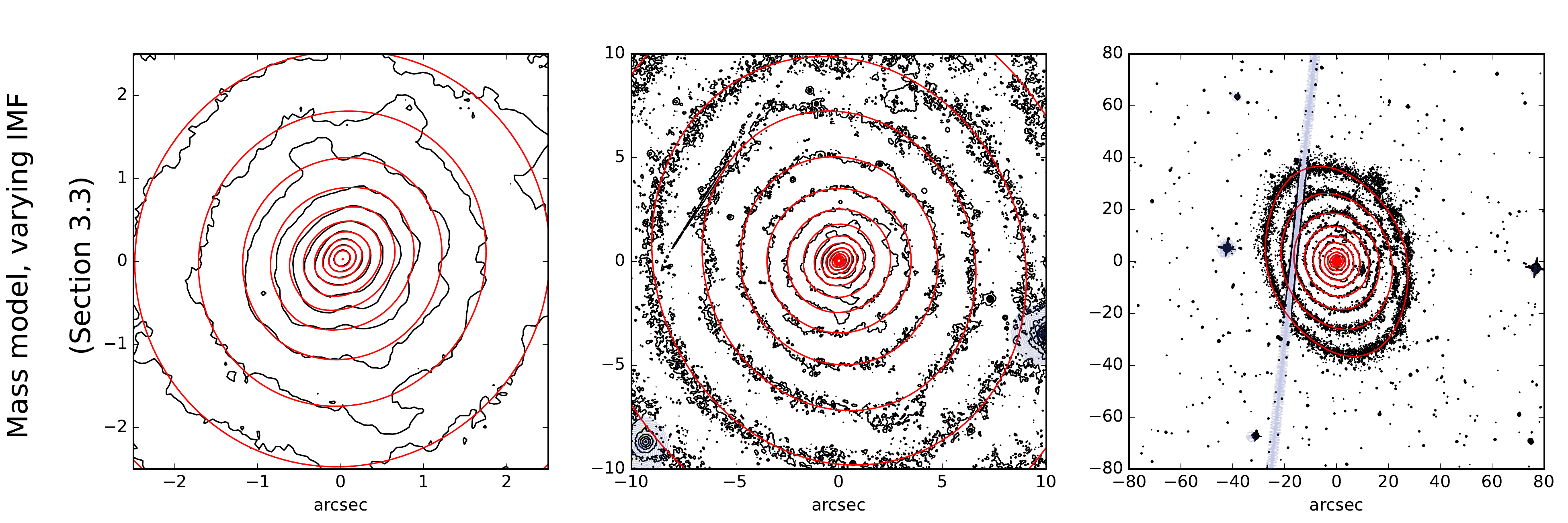}

      \caption{Overview of the light and mass models used in this paper. Top: Isophotes of the HST/ACS F850LP band imaging data of FCC\,47 (black) overplotted with the MGE model from Section 3.1 (red). Middle: Mass map derived by multiplying the luminosity in the HST/ACS F850LP image with the $M_{\rm *,MW-IMF}/L$ (fixed IMF) shown in the left panels of Fig. \ref{ff:ml}. Overplotted is the MGE model from Section 3.2 (red). Bottom: Mass map derived by multiplying the luminosity in the HST/ACS F850LP image with the $M_{\rm *,free-IMF}/L$ (varying IMF) shown in the right panels of Fig. \ref{ff:ml}. Overplotted is the MGE model from Section 3.3 (red). From left to right, we show a cut-out of the central $5\times 5$ arcsec that is dominated by the NSC, a zoom out to $20''\times 20''$ , and a zoom out to $160''\times 160''$ covering the full galaxy. Masked regions are shaded in grey. Only for this visualisation, the ACS image and the MGE were rotated such that North is aligned with the y-axis.
      }
      \label{ff:mge}
\end{figure*}

\subsection{Constant mass-to-light ratio}

We created the first mass model of FCC\,47 by applying the Multi-Gaussian Expansion (MGE) method \citep{Emsellem1994,Cappellari2002} to the ACS F850LP image of FCC\,47. The NSC in FCC\,47 is photometrically decoupled from the host galaxy with a photometric twist of about 90 degrees. We therefore decided to have a varying position angle for the different Gaussian components. This can be achieved by using the \verb@mge_fit_sectors_twist@ routine. 

As the ACS F850LP image has both high resolution and large coverage, it is sufficient to create the MGE fit on this image alone. During the fit, we masked stars and the gap between the two ACS chips. We also took the point spread function (PSF) of the ACS image into account, which we generated using the TinyTim PSF modelling tool \citep{Krist2001}. The PSF was corrected for distortion as explained in \cite{Thater2019} and parametrised as a sum of Gaussians. The PSF parametrisation is shown in Table \ref{tt:hst_psf} of the Appendix.

In the top panels of Figure \ref{ff:mge}, we show the contours of our final MGE overplotted on the ACS imaging data. Our MGE reproduces the regular isophotes of the galaxy very well. The images clearly indicate that FCC 47 shows no dust features that would require an extended galaxy dust-correction. Also, the strong photometric twist between the centre and the host galaxy is well recovered by our MGE. 

In total, we obtained 14 MGE components that are listed in Table \ref{tt:mge} in the Appendix. Each Gaussian is assigned an observed axis ratio  $q_j$ and a misalignment angle $\Psi_j$. For the photometric calibration, we followed the standard conversion to AB magnitude system using the zero points from \cite{Sirianni2005} and assuming 4.50 mag for the absolute magnitude of the Sun in the F850LP band \citep{Willmer2018} and a Galactic extinction of AF160 = 0.014 mag \citep{Schlafly2011}. 

The two-dimensional light parametrisation is then (assuming a triaxial potential and the intrinsic shape of the galaxy) deprojected into three-dimensional space. By multiplying by the constant $M_{\rm *,const}/L$ of the dynamical models, we obtained a model of the mass density from which the gravitational potential is calculated by solving Poisson's equation.

\begin{figure}
  \centering
      \includegraphics[width=0.49\textwidth]{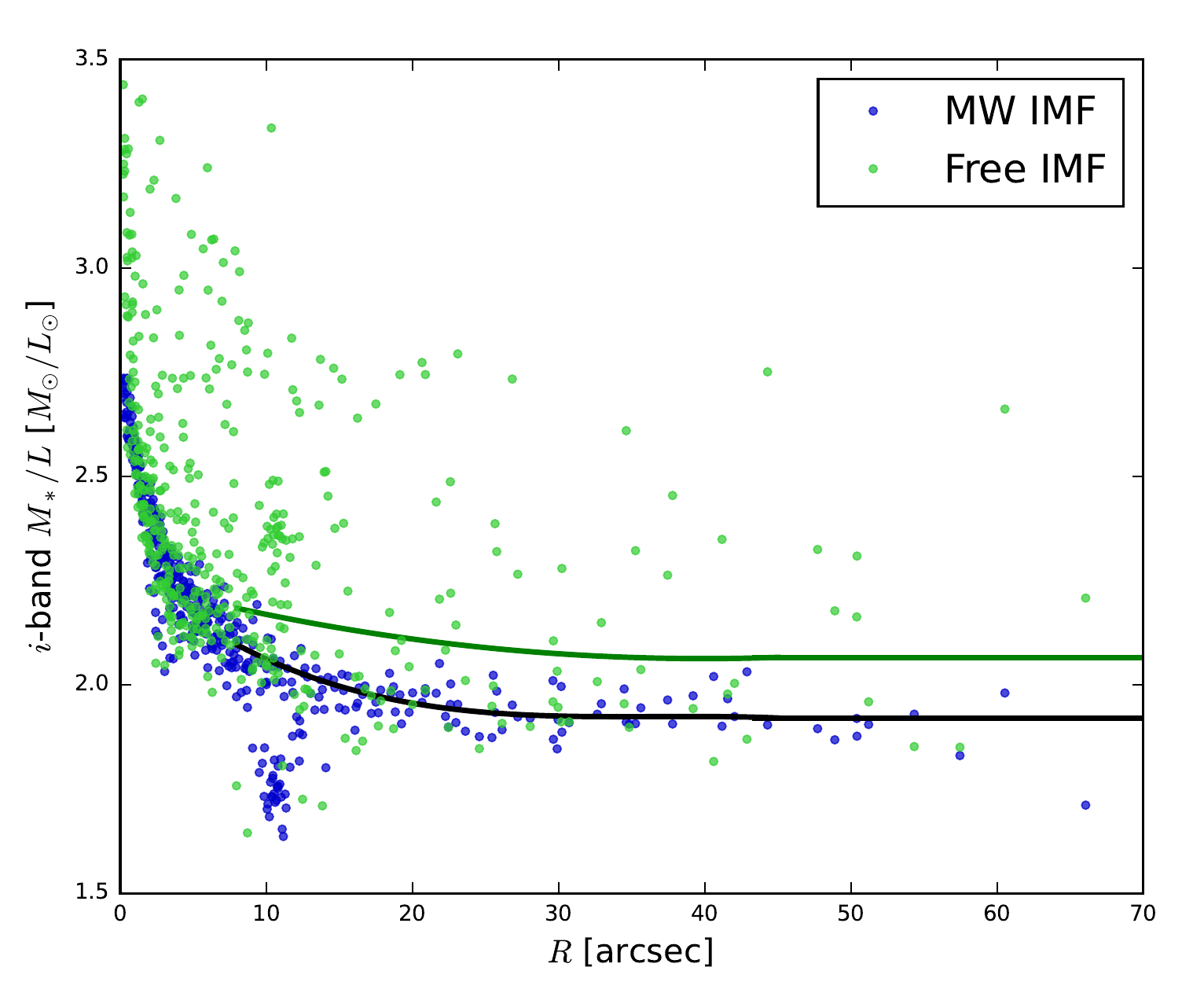}

      \caption{Stellar mass-to-light ratio ($M_*/L$) profiles in the i-band. Each point is a Voronoi bin as shown in Fig. 2. The blue points correspond to the left maps in Fig. 2, while the green points correspond to the maps on the right. We combined our $M_*/L$ maps with the HST image to derive mass maps. As the HST image has a larger coverage than the MUSE data, we had to partially interpolate the $M_*/L$ values. The lines show the interpolated values used outside of the MUSE coverage ($R>8$ arcsec; see explanation in Section 3.2 and Fig. C.1).  
      }
      \label{ff:ml_gradient}
\end{figure}

\subsection{Varying mass-to-light ratio with fixed MW-like IMF}
We ran a second MGE of FCC\,47 on a \textit{mass map} of the galaxy. The mass map was derived by multiplying the above-mentioned ACS image with the varying stellar mass-to-light ratio ($M_{\rm *,MW-IMF}/L$) in the i-band. The $M_{\rm *,MW-IMF}/L$ map was derived via stellar population analysis as follows. We used pPXF and fitted the MUSE data of FCC\,47 between 4700 and 7100\AA\, with the E-MILES single stellar population (SSP) templates \citep{Vazdekis2010, Vazdekis2016}. The wavelength region was chosen to avoid the regions at longer wavelengths that can be affected by sky residuals. The E-MILES SSP templates sample a grid of single stellar populations with ages ranging from 30 Myr to 14 Gyr and total metallicities between [M/H] = $-2.27$ dex and $+0.4$ dex, at a spectral resolution of 2.5\AA. We used the description of the MUSE line spread function from \cite{Guerou2016}.

To sample random uncertainties, each bin was fitted ten times with a random initialisation of the spectrum (see e.g. \citealt{Cappellari2004}). No regularisation was applied as we are only interested in the mean age and metallicity of each bin. To obtain mass-to-light ratios, we used the E-MILES photometric predictions\footnote{\url{http://research.iac.es/proyecto/miles/pages/photometric-predictions-based-on-e-miles-seds.php}} that associate stellar mass-to-light ratios in different bands with each single stellar population template. Using a simple grid interpolation, the $M_{\rm *,MW-IMF}/L$ in the $i$-band was derived. The variations in $M_{\rm *,MW-IMF}/L$ (shown in the left panel of Fig. \ref{ff:ml}) are mostly driven by variations in the metallicities as the ages are relatively uniform at 12 Gyr and older (see Figure 7 in \citealt{Fahrion2019}). 

This stellar $M_{\rm *,MW-IMF}/L$ - map was then multiplied with the imaging data (in units of $L_{\odot}$) to obtain a map of the stellar mass. It is important to understand that there are two limitations to this approach. First, the spatial resolution of the MUSE data ($\approx 0.7\arcsec$ FWHM) is not as high as for the HST imaging ($\approx 0.09\arcsec$ FWHM). Figures \ref{ff:ml} and \ref{ff:ml_gradient} clearly show that the ($M_{\rm *,MW-IMF}/L$) map has an almost constant value of 2.74 $M_{\odot}/L_{\odot}$ within the central 1 arcsec. Moreover, the true $M_{\rm *}/L$ is likely higher in the unresolved centre of FCC 47, and we slightly underestimate the stellar mass here. However, our approach is more accurate than when assuming a constant value $M_{\rm *,const}/L$ (that is lower than 2.74 $M_{\odot}/L_{\odot}$). Second, the FOV coverage of the  MUSE data is much smaller than for the ACS image and also only covers one quadrant of the galaxy. We extrapolated the $M_*/L$ to the galaxy quadrant not covered by the MUSE FOV by calculating the average $M_{\rm *,MW-IMF}/L$ at each elliptical distance from the centre and assigning this average value to ACS data points outside of the MUSE coverage. For r > 50 arcsecond, we assigned a constant $M_*/L$ of 1.92 $M_{\odot}/L_{\odot}$, the average value at the largest radii covered with MUSE. The $M_{\rm *,MW-IMF}/L$ profiles in Figure \ref{ff:ml_gradient} show that for $R > 30''$ the $M_*/L$ stays constant and we therefore also do not expect a significant change for larger radii.

In order to derive a mass map, the $M_{\rm *,MW-IMF}/L$ map is multiplied to the ACS luminosity map pixel by pixel. Consequently, we rotated the MUSE $M_{\rm *,MW-IMF}/L$- map by 98$^{\circ}$ to match the orientation of the ACS imaging data and re-sampled the $M_{\rm *,MW-IMF}/L$ bins to the ACS pixel size of $0.05''$. We present the derived $M_{\rm *,MW-IMF}/L$ map in Fig. C.1 (left panel) of the Appendix. The changes in $M_{\rm *,MW-IMF}/L$ are very smooth as the stellar populations change only slightly. After multiplication with the ACS luminosity map, we obtained a detailed mass map of FCC\,47 accounting for variation in the stellar population. 

This mass map was then fitted with MGE in an analogous way to the previous section. We masked the stars and the gap between the two ACS chips and took the ACS PSF into account. In the middle row of Figure \ref{ff:mge}, we present the contours of our final MGE overplotted on the mass map. A comparison with the light model shows that the lines of constant mass have a very similar shape as the isophotes. The MGE is able to easily depict the mass model. It should be noted that the mass model obtained with $M_{\rm *,MW-IMF}/L$ decreases faster in the outskirts than the mass model obtained with $M_{\rm *,const}/L$, because the $M_*/L$ is lower in the outskirts than in the central regions. In total, we obtained the 14 MGE mass components listed in Table \ref{tt:mge2} in the Appendix.

\begin{figure}
  \centering
      \includegraphics[width=0.49\textwidth]{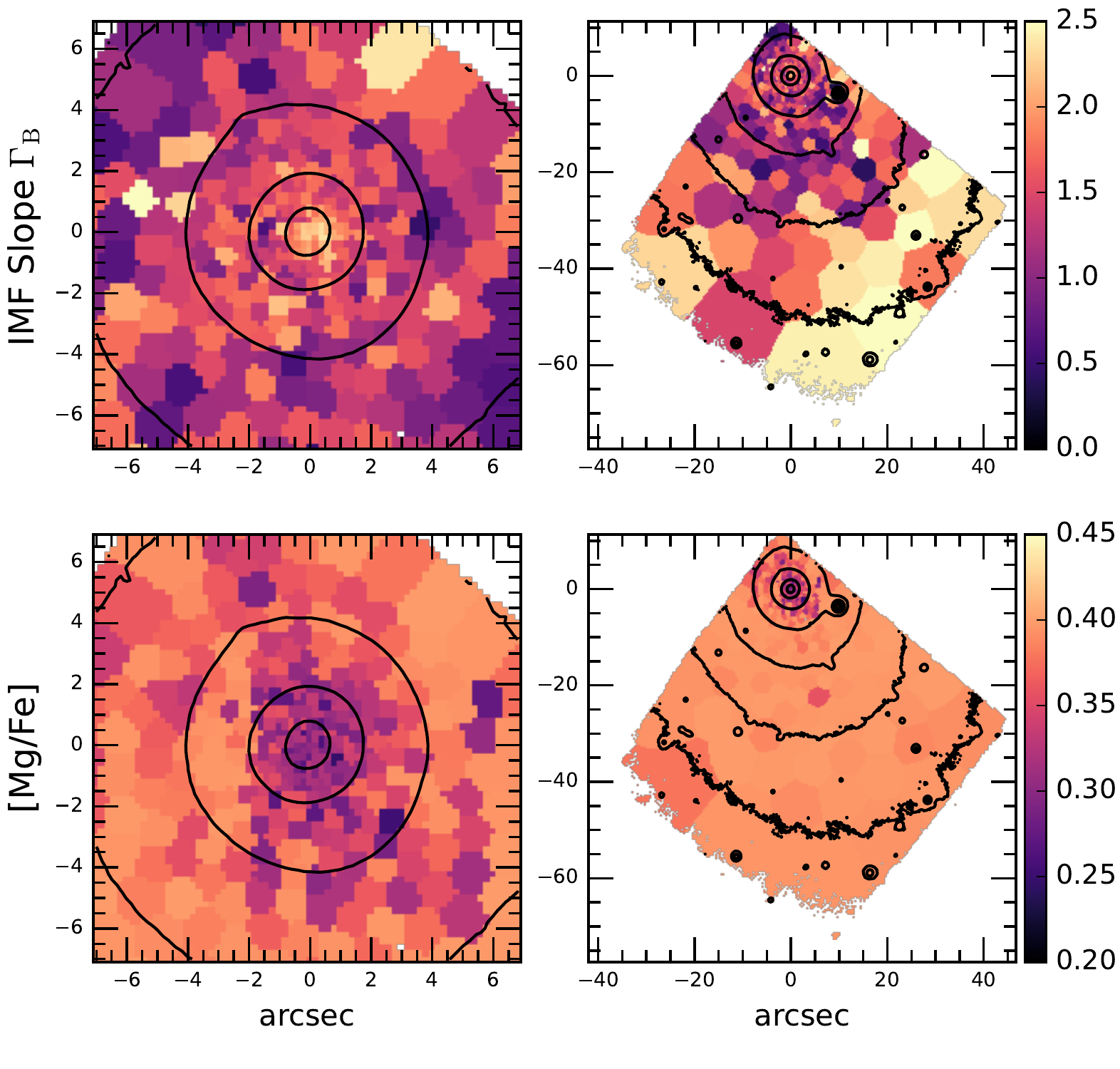}

      \caption{[Mg/Fe] and IMF slope maps $\Gamma_{\rm B}$ of FCC 47. Left and right: Maps extracted from the MUSE observations shown within a zoom onto the centre and across the full MUSE FoV. The NSC of FCC 47 has a larger IMF slope and lower alpha-element abundance.
      }
      \label{ff:imf}
\end{figure}

\subsection{Varying mass-to-light ratio with varying IMF}
In the third approach, we created a \textit{mass map} of FCC 47 by multiplying a different varying mass-to-light ratio map with the ACS image. The mass-to-light ratio map with a spatially varying IMF ($M_{\rm *,free-IMF}/L$) was derived following our previous experience with similar MUSE data described in \cite{Martin-Navarro2019} and \cite{Martin-Navarro2021}. Briefly, our approach to measure IMF variations in the FCC47 MUSE data consists of two steps. First, we derive the mean age of each (Voronoi-binned) spectrum using pPXF. There are two main differences between this approach and the age determination described in the previous section. First, we regularised the pPXF solution in order to minimise bin-to-bin age variations that would propagate to the second step. In addition, we also allowed for a variable IMF while measuring the best-fitting luminosity-weighted age from pPXF. After this first step, we used the derived ages to measure the metallicity, [Mg/Fe], [Ti/Fe], and IMF slope $\Gamma_{\rm B}$ of each bin with the so-called full-index fitting method (see \citealt{Martin-Navarro2019}). In practice, this method is a hybrid approach combining the classical line-strength analysis with the advantages of the full spectral fitting. This two-step IMF-fitting process was fed with the MILES alpha-variable stellar population models \citep{Vazdekis2016}. Maps of the best-fit IMF slope $\Gamma_{\rm B}$ and [Mg/Fe] are shown in Figure \ref{ff:imf}. Spatial variations in the IMF slope are clearly visible with higher slopes towards the centre of FCC 47. We used the best-fitting age, metallicity, and IMF slope to derive the $M_{\rm *,free-IMF}/L$ map in the i-band.

Using the free IMF, we noticed very high stellar $M_*/L$ in some bins, reaching up to about 8 $M_{\odot}/L_{\odot}$ or more. High stellar $M_*/L$ values coincided with very low IMF values. These strong $M_*/L$ variations are entirely due to the mass of stellar remnants. However, from the point of view of the stellar population analysis, we are completely blind to these remnants (they do not contribute to the light budget) and therefore the abnormally high $M_*/L$ ratios are just reflecting our lack of observational constraints on the high-mass and very low-mass ends of the IMF (i.e. they are likely just a modelling artifact). Furthermore, these strong variations were so strong that it was not possible to create a smooth MGE. We therefore ran the routine for a second time, disallowing IMF slopes lower than 0.8. This change left the centre of the $M_{\rm *,free-IMF}/L$ map unchanged. The resulting $M_{\rm *,free-IMF}/L$ map in the i-band is shown in the right panels of Figure \ref{ff:ml}. It should be noted that the $M_{\rm *,free-IMF}/L$ values are about 0.2 $M_{\odot}/L_{\odot}$ higher than the $M_{\rm *,MW-IMF}/L$ map from SSP models with a MW-like IMF (See Figure \ref{ff:ml_gradient}). Furthermore, the maps are more centrally peaked when taking into account the IMF. In the centre of the galaxy, the stellar $M_{\rm *,free-IMF}/L$ reaches 3.3 $M_{\odot}/L_{\odot}$. 

We followed the same procedure as in Section 3.2 to extrapolate the $M_{\rm *,free-IMF}/L$ in the outskirts and resample and rotate it to resemble the ACS image. The resulting interpolated $M_*/L$ map is presented in Fig. C.1 (right panel) of the Appendix. We also fitted an MGE to this mass map using the same setup as in the previous sections. We masked the stars and the gap between the two ACS chips and took the ACS PSF into account. In the bottom panels of Figure \ref{ff:mge}, we present the contours of our final MGE overplotted on the mass map. In general, the MGE can nicely recover the shape of this mass map. However, due to the stronger variations in $M_*/L$, the mass map shows many jagged features that cannot be recovered by a smooth MGE. It should also be noted that the spatial extent of the galaxy in this mass model is slightly smaller than for the mass model in Section 3.2 due to a steeper $M_*/L$ gradient (low $M_*/L$ in the galaxy outskirts). In total, we obtained 14 MGE mass components that are listed in Table \ref{tt:mge3} in the Appendix.

\section{Dynamical models}
\label{s:dyna}

\subsection{Settings}
 The dynamics of the NSC and its host galaxy FCC\,47 were modelled by \cite{Fahrion2019} using large-scale MUSE observations. We extended their analysis in this study, using high-resolution HST imaging data and SINFONI kinematics to include detailed information of the nucleus of FCC\,47 in our dynamical modelling. We furthermore took the variablity of the stellar $M_*/L$ from both stellar populations and IMF variation into account. In order to model FCC\,47, we used the DYnamics, Age, and Metallicity Indicators Tracing Evolution \citep[DYNAMITE v2.0;\footnote{\url{https://www.univie.ac.at/dynamics/dynamite_docs/index.html}}][]{Jethwa2020,Thater2022b} software, which is a publicly available updated version of the triaxial Schwarzschild orbit-superposition code developed by \cite{vandenBosch2008}.
 
\begin{figure}
  \centering
      \includegraphics[width=0.5\textwidth]{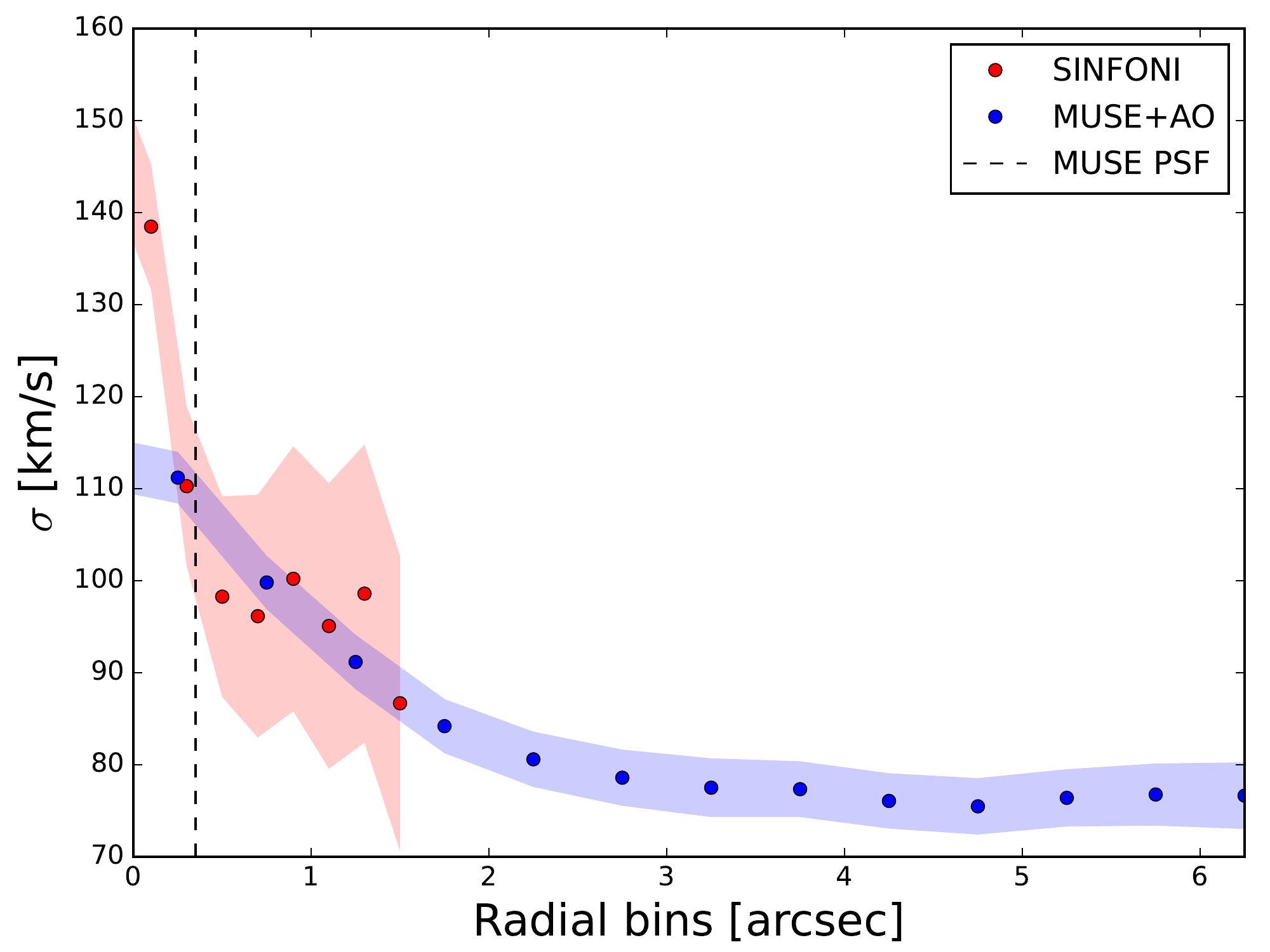}

      \caption{Radial comparison of the velocity dispersion derived for the two different IFU data sets. The values were averaged within circular annuli around the kinematic centre. The error range of the averaged values in the radial bins are calculated via error propagation and are shown as shaded regions. SINFONI and MUSE maps match well for radii larger than the MUSE PSF ($0.7\arcsec$ FWHM).
      }
      \label{ff:kin_comp}
\end{figure} 
 
 The triaxial Schwarzschild orbit-superposition technique works as follows. Based on the assumed gravitational potential of the galaxy, an orbit library is calculated and then fitted to the observations. Our assumed gravitational potential consists of a stellar component, a central black hole, and dark matter parametrised as spherical halo with Navarro-Frenk-White \citep[NFW; ][]{Navarro1996} radial profile. The stellar potential is derived from the MGE (Section \ref{s:mge}) by deprojecting the two-dimensional surface brightness into a three-dimensional luminosity density using the triaxial viewing angles ($\theta,\phi,\Psi$) of the galaxy. Here, $\theta$ corresponds to the classical inclination of the galaxy. The luminosity density is then turned into a mass density assuming a constant or varying $M_*/L$.
 
 The viewing angles are directly connected to the intrinsic shape parameters of the galaxy: $p$ (intrinsic intermediate-to-major axis ratio), $q$ (intrinsic minor-to-major axis ratio), and $u$ (ratio between projected and intrinsic major axis). The detailed conversion is given in equations (7) to (9) of \cite{vandenBosch2008} and equations (16) to (21) of \cite{Jin2019}. 
The black hole is added via a Plummer potential, which is described with the black hole mass $M_{\rm BH}$ and a small but non-zero softening length (that is not varied during the fit). The NFW dark matter profile is parametrised by $f_{\rm DM}=M_{200}/M_*$, the dark mass fraction within the virial radius $r_{200}$, and the concentration parameter $c = R_{200}/R_{\rm s}$, where $R_{\rm s}$ is the scale radius of the NFW profile.

For each triaxial gravitational potential described by the parameters $p, q, u $, $M_{\rm BH}$, $f_{\rm DM}=(M_{200}/M_{*})$, and $c$, we generated two representative orbit libraries: one orbit library that mostly represents long- and short-axis tube orbits and a second orbit library for box orbits. The tube orbits are characterised by three integrals of motion: the binding energy E, the second integral I2, and third integral I3. Our box orbits are described by the binding energy E and two spherical angles ($\theta,\phi$). We sampled the tube orbit libraries with $35 \times 11 \times 11$ combinations of (E, I2, I3) including prograde and retrograde rotation. For the box orbits, we also created $35 \times 11 \times 11$ combinations of (E, $\theta,\phi$). In addition, we dithered each orbit by slightly perturbing the initial conditions in order to smooth the orbit distribution. The orbit trajectories created by the dithering are co-added to form an orbit bundle in our
orbit library. Each resulting orbit bundle contains $5^3=125$ orbits. 

Our Schwarzschild orbit-superposition code derives the weights of each orbit using the \cite{Lawson1974} non-negative least square (NNLS) implementation. Therefore,
model kinematics were generated from the orbit libraries and compared to the observed MUSE and SINFONI kinematics simultaneously, while also taking their PSF into account. The high-resolution SINFONI kinematics were used to constrain the dynamics in the centre of the galaxy. As the kinematic uncertainties of the MUSE observations are a factor of three lower than the kinematic uncertainties of the SINFONI observations (see Figure \ref{ff:kin_comp}), and therefore bias the dynamical models, we had to mask the central 1 arcsec of the MUSE kinematics. In addition, we masked the bright foreground star close to the galaxy centre and the ultracompact dwarf to the southeast of FCC\,47 that was discovered by \cite{Fahrion2019b}. 

A significant advantage of the numerical computation is that a change in $M_*/L$ does not change the shape of the gravitational potential. We are therefore not required to run additional orbit libraries when changing $M_*/L$ in the Schwarzschild model, but need to only up- or downscale the velocities in the existing orbit libraries. Furthermore, when assuming a {varying} $M_*/L$ from variation in age, metallicity, or IMF, the MGE describes the mass density of the stellar component and we only need a mass scaling factor $S$ in the dynamical models. This mass scaling factor S is typically close to unity, but can slightly differ due to an inaccurate parametrisation of dark matter, incorrect normalisation of the SSP estimates of the $M_*/L,$ or variation of the IMF. 
In order to not confuse the reader with changing parameters we use the mass scaling factor $S$ in the following sections. We emphasise that for the dynamical model with $M_{\rm *,const}/L,$ the scaling factor S corresponds to the commonly used $M_*/L$ in F850LP band.
This means that each calculated gravitational potential depends on seven parameters, which are varied in order to find the best-fitting model, namely $p, q, u , S$, $M_{\rm BH}$, $f_{\rm DM}=(M_{200}/M_{*})$, and $c$.

\begin{figure*}
  \centering
      \includegraphics[width=0.49\textwidth]{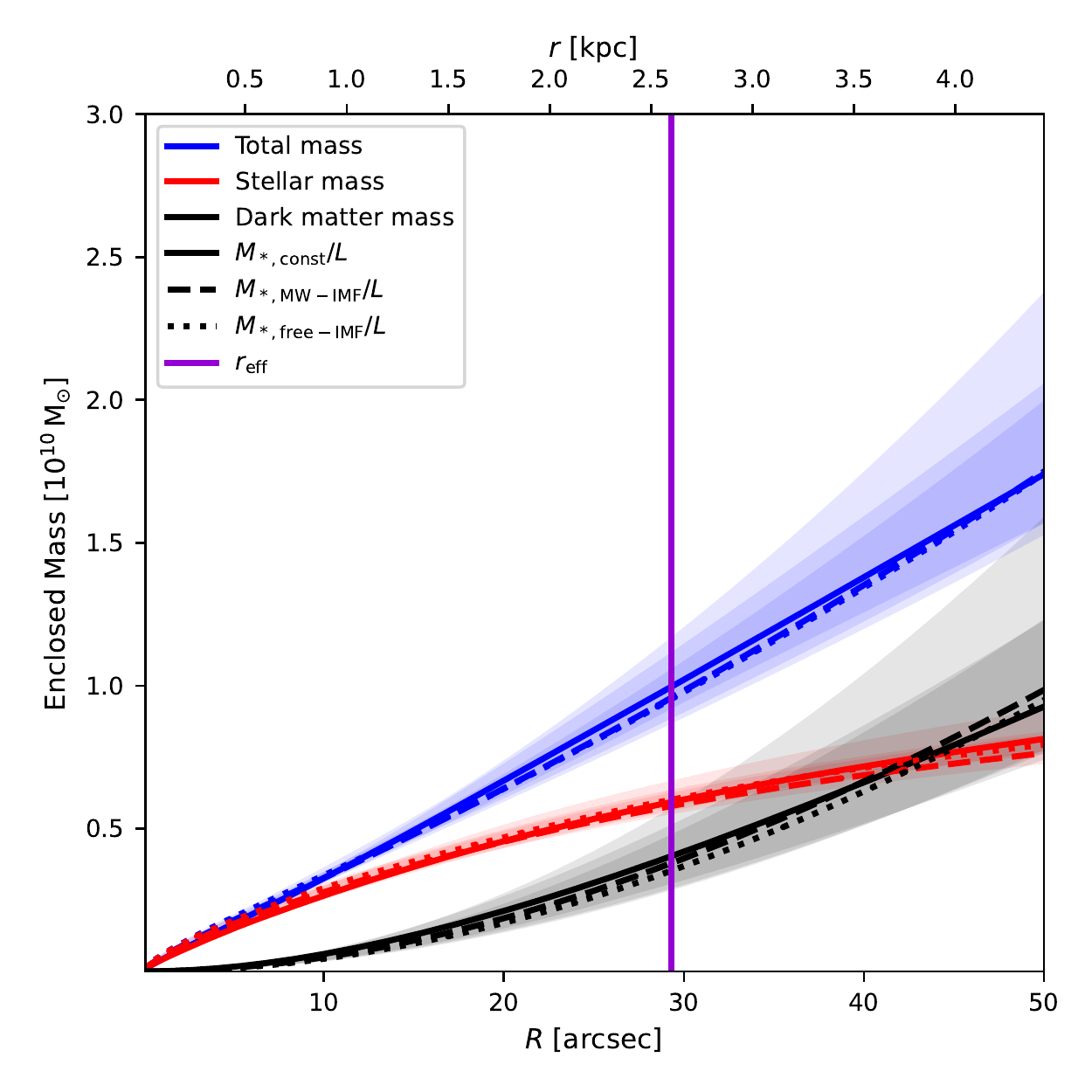}
 \includegraphics[width=0.49\textwidth]{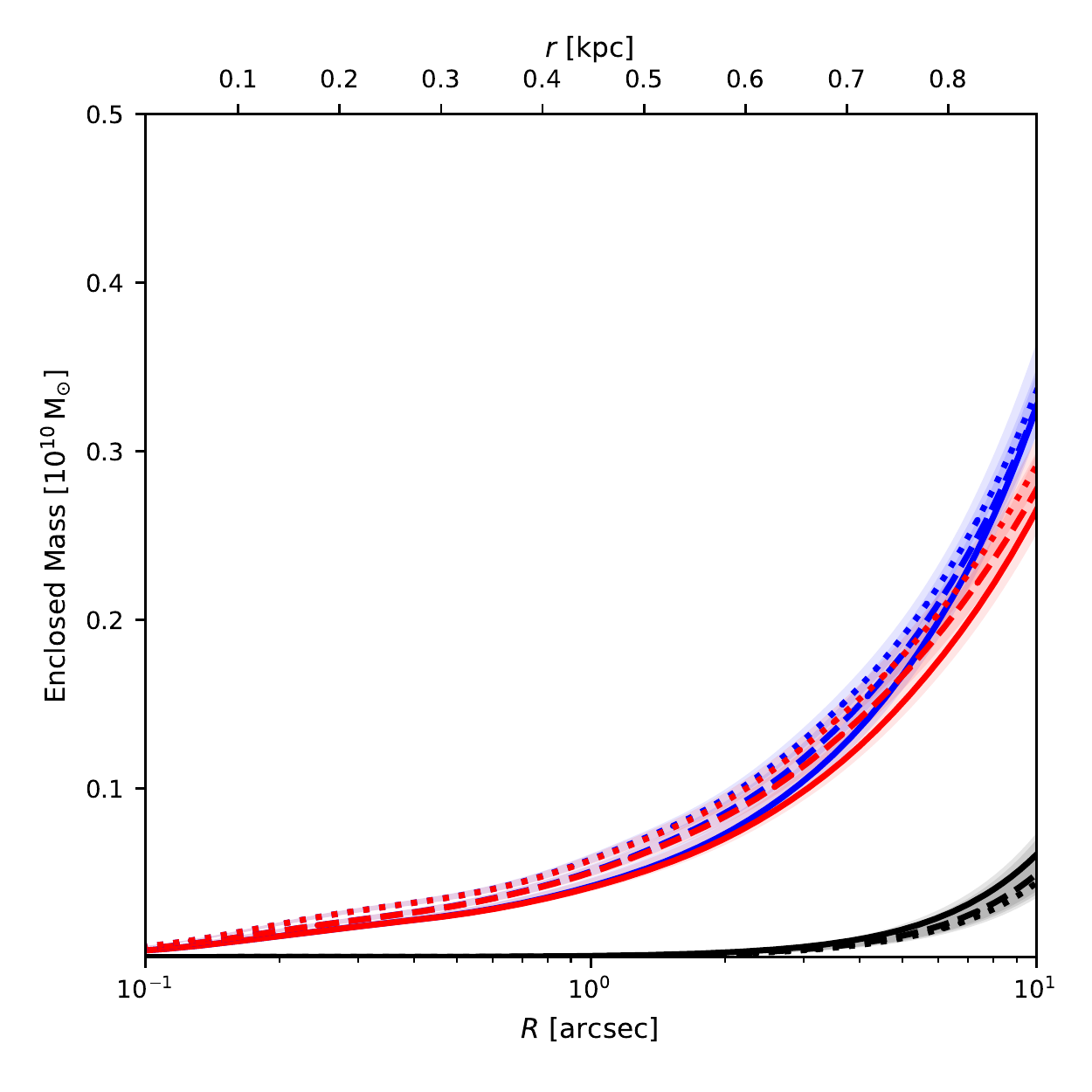}
      \caption{Enclosed mass profiles for the stellar component (red), dark matter component (black), and their combination (blue) for the different DYNAMITE modelling runs of FCC\,47. The left panel shows the profiles over the full FoV of the MUSE observations and the right panel only shows the central 10 arcsec in logarithmic scales. The central mass profile is clearly dominated by stellar mass. The shaded regions indicate the $1\sigma$ uncertainties of the mass profiles. The spatial resolution of the SINFONI observations is $0.14\arcsec$ (FWHM). The purple line visualises the effective radius of FCC 47.}
      \label{ff:enclosed_mass}
\end{figure*}

\begin{figure*}
  \centering
      \includegraphics[width=1.0\textwidth]{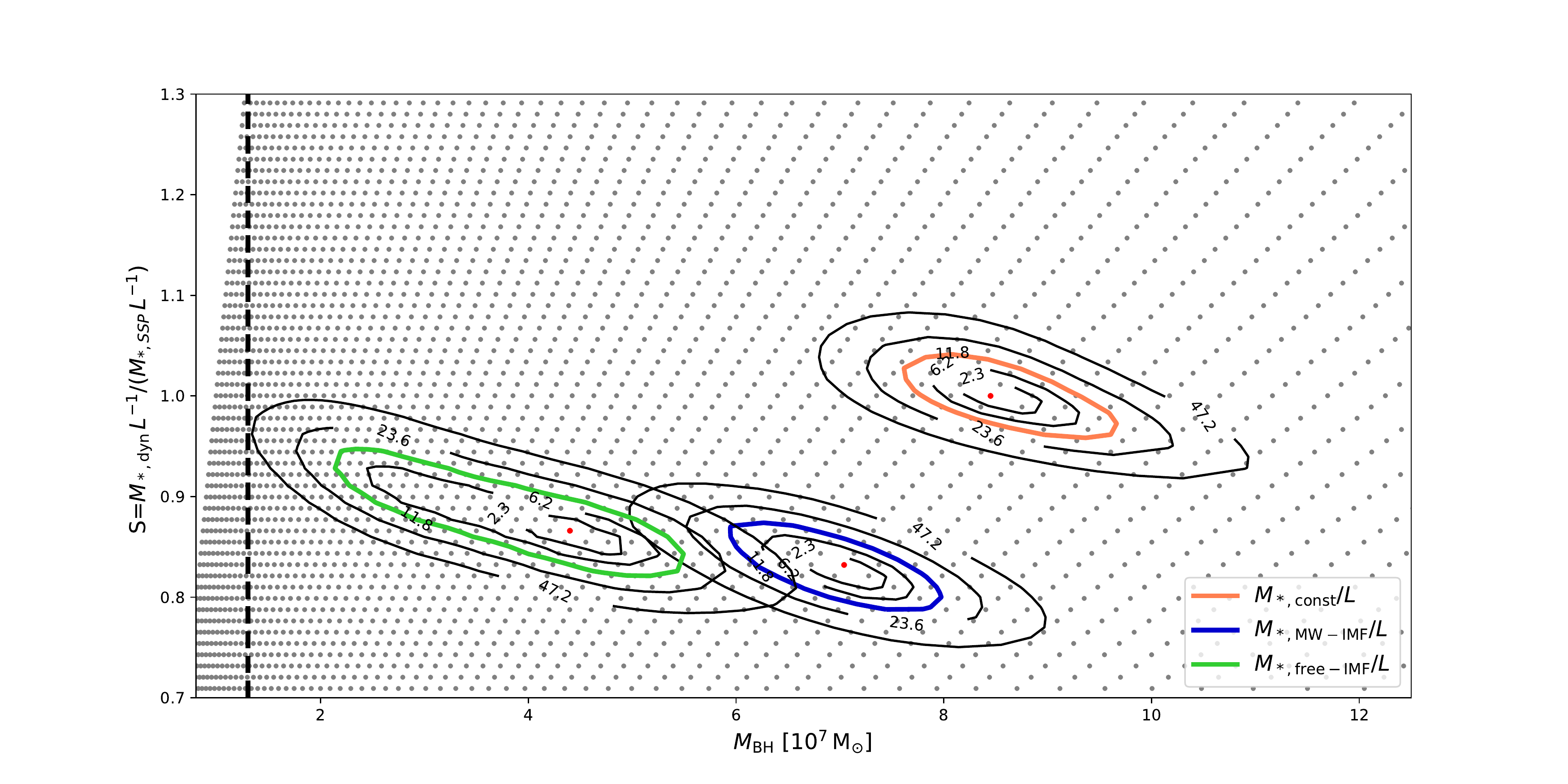}

      \caption{Results of the triaxial Schwarzschild modelling performed with DYNAMITE. After constraining the intrinsic shape and dark matter parameters, a fine grid of models was run to derive the mass scaling factor S and the central black hole mass $M_{\rm BH}$. S accounts for differences in the stellar M/L derived from SSP modelling and dynamical modelling. Each Schwarzschild model is indicated as a grey dot. The overplotted contours indicate the $\Delta \chi^2 = \chi^2-\chi^2_{min}$ levels. We show the results from all three modelling runs with different assumptions on the stellar mass-to-light ratio $M_*/L$ in this figure. The coloured contours indicate the $3\sigma$ confidence intervals for the different modelling runs. The mass measurements are robust as indicated by the closed $\chi^2$-contours outside of the $3\sigma$ confidence interval. The best-fitting models were derived as the minimum of the $\chi^2$ distribution and are indicated with a red dot, respectively. For the models with constant stellar M/L, we normalised the dynamical stellar M/L by the best-fit M/L.The vertical black line indicates the smallest black hole mass that we expect to robustly detect based on the SoI argument and the spatial resolution of the SINFONI observations.}
      
      \label{ff:black_hole_grid}
\end{figure*}

\begin{figure*}
  \centering
      \includegraphics[width=0.75\textwidth]{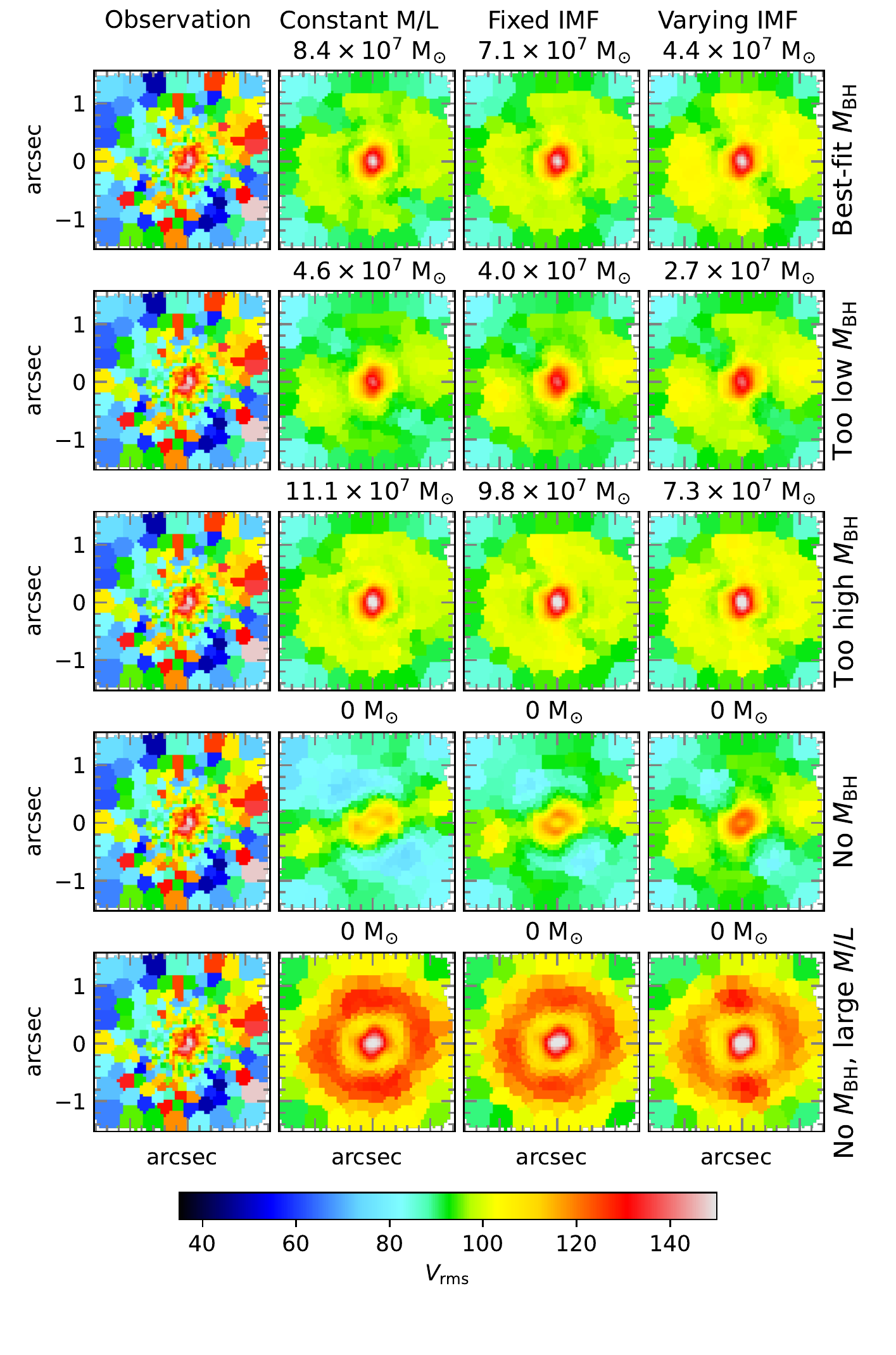}

      \caption{Visual comparison between the observed SINFONI $V_{\rm rms} = \sqrt{V^2+ \sigma^2}$ map and the $V_{\rm rms}$ maps of the DYNAMITE triaxial Schwarzschild models with best-fitting $M_{\rm BH}$, excessively high and excessively low black hole mass, and a non-existent black hole at the best-fitting mass scaling factor S for each set of models. The two right-most columns are models calculated with varying $M_*/L$. North is up and east to the left. 
      }
      \label{ff:bh_comparison}
\end{figure*}

\subsection{Intrinsic shape and dark matter parameters}
\label{ss:global_results}
We started our orbit-superposition models by first constraining the intrinsic shape and dark matter parameters for each of the three different mass models from Section 2. For our models with $M_{\rm *,const}/L$, we started our parameter search centred on the dynamical results of \cite{Fahrion2019} and used an iterative approach as described in \cite{Zhu2018a,Zhu2018b} and \cite{Santucci2022}. Our starting parameters were $p=0.60, q=0.25,  u=0.98, \log f_{\rm DM}= 1.25,$  $\log c =1.40,$ and $S=2.50$ M$_{\odot}/$L$_{\rm \odot}$. After each iteration, the best-fit model was derived. Around this best-fit model, we calculated new models at a predefined step size. The step size was adapted to the existing models: Large initial step sizes to avoid local minima and smaller step sizes around the best-fit model. In total, we ran 8000 models for the mass model with $M_{\rm *,const}/L$. Figure \ref{ff:chi2_global} in the Appendix shows the $\chi^2$ distribution of our final model grid assuming $M_{\rm *,const}/L$ . From the minimum of the  $\chi^2$ distribution, we estimated the shape parameters of $p = 0.59 \pm 0.03 $, $q = 0.23 \pm 0.06,$ and $u = 0.93\pm 0.03 $ (corresponding to viewing angles $\theta=47^{\circ}$, $\phi=56^{\circ}$ and $\psi=110^{\circ}$), a NFW dark matter profile with $\log f_{\rm DM}= 1.25 \pm 0.30 $ and $\log c=1.19 \pm 0.20 $, and $S$ of $1.80 \pm 0.20$ $M_{\odot}/L_{\odot}$ (in the i-band) at $3\sigma$ confidence level. Due to the choice of putting FCC\,47 in the upper corner of the MUSE FoV, we cover almost two effective radii and can constrain the dark matter NFW parameters of FCC\,47 remarkably well. Our derived best-fit values are consistent with those obtained by \cite{Fahrion2019}. Small differences are expected as we probe different galaxy scales and use different kinematics and photometric data. 

Our derived mass scaling factor S is lower than the global $M_*/L_{\rm r}$ of \cite{Fahrion2019}, because they used r-band VST data while we use i-band HST/ACS data to derive the galaxy surface brightness model. We visually inspected the stellar $M_*/L_{\rm r}$ map derived from SSP synthesis of the MUSE data and noted that it was consistent with the $M_*/L_{\rm r}$ of \cite{Fahrion2019}.

When running Schwarzschild models on the two mass models with spatially varying $M_*/L$, we obtained similar global parameters as for the models with $M_{\rm *,const}/L$. The $\chi^2$ distributions of the models using $M_{\rm *,MW-IMF}/L$ and $M_{\rm *,free-IMF}/L$ are shown in Figures \ref{ff:chi2_global2} and \ref{ff:chi2_global3} in the Appendix. For $M_{\rm *,MW-IMF}/L$, the best-fit dark matter fraction is marginally higher due to the lower stellar mass in the outskirts of the galaxy. While the models with $M_{\rm *,const}/L$ predict a S of 1.80 $M_{\odot}/L_{\odot}$ in the outskirts of the galaxy, this value decreases to $M_{\rm *,MW-IMF}/L$ $\times$ S = $1.92 \, $M$_{\odot}/L_{\odot}\times 0.83$ = 1.59 $M_{\odot}/L_{\odot}$ using a spatially varying mass-to-light ratio. The dynamical models compensate this loss in stellar mass by slightly increasing the dark matter fraction (see Figure \ref{ff:enclosed_mass}). On the other hand, for $M_{\rm *,free-IMF}/L$ the effective value for S is slightly higher than for $M_{\rm *,const}/L,$ which leads to mildly lower dark matter fraction. However, the changes in the enclosed mass of dark matter are much smaller than the $1\sigma$ uncertainties, and the derived black hole masses are not significantly affected \citep{Thater2022}. All global best-fit parameters are listed in Table \ref{tt:results}.

  \begin{table*}
  \label{tt:results_grid}
\caption{Results of the regular grid search for Schwarzschild models with different stellar mass models.}
\centering
\begin{tabular}{lccc}
\hline\hline
Mass model  & $M_{\rm BH}$  &  $S$  & $\chi^2$/d.o.f.\\
     & ($10^7\,$M$_{\odot}$)    \\
 (1) & (2) & (3) & (4)    \\
\hline
Constant $M_*/L$  ($M_{\rm *,const}/L$) & 8.4$^{+1.3}_{-0.9}$ & *1.82$\pm 0.07$ & 2.14 \\
Varying $M_*/L$ \& fixed IMF ($M_{\rm *,MW-IMF}/L$) & 7.1$^{+0.8}_{-1.1}$ & 0.83$\pm 0.04$  & 2.15\\
Varying $M_*/L$ \& varying IMF ($M_{\rm *,free-IMF}/L$) & 4.4$^{+1.2}_{-2.1}$ & 0.86$\pm 0.04$ & 2.18\\
\hline
\end{tabular}
\\
\tablefoot{This table summarises the results of Section 4.3. Column 1: Details of the stellar mass model used for the dynamical modelling. Columns 2 and 3: Parameters of the best-fitting models derived from the regular grid search: black hole mass $M_{\rm BH}$ and scaling factor S.  
Column 4:  $\chi^2$ over the degrees of freedom. The uncertainties were calculated using $3\sigma$ confidence intervals.
*For the model with constant mass-to-light ratio, the mass scaling factor S corresponds to the derived stellar $M_*/L_{\rm F850}$ in M$_{\odot}/$L$_{\rm \odot}$.}
\end{table*}

\subsection{Central black hole}

After having found the global best-fit parameters for the gravitational potential of the galaxy, we ran a second very fine grid to derive a robust estimate of the central black hole mass for the three different mass models from Section 3. We kept the original data input, fixed the global parameters of the gravitational potential to the best-fit values shown in Table \ref{tt:results} in the Appendix, and only varied the mass scaling factor S and $\log (M_{\rm BH}/M_{\odot})$ in steps of 0.02 and 0.02 dex, respectively.

Figure \ref{ff:black_hole_grid} shows the final grid to estimate the central black hole mass. Each of the black dots indicates a model. Plotted on the grid are the $\chi^2$ distributions as a function of $M_{\rm BH}$ and mass scaling factor S. In order to smooth the topology of the $\chi^2$ contours, we applied the local regression smoothing algorithm LOESS \cite{Cleveland1979} as adapted for two dimensions by \cite{Cappellari2013}. We show the $\chi^2$ distributions (for clarity up to $\Delta \chi^2 = \chi^2-\chi^2_{min}= 47.2$) overplotted as contours: in orange the $\chi^2$ distribution of the mass model assuming $M_{\rm *,const}/L$, blue assuming $M_{\rm *,MW-IMF}/L,$ and green $M_{\rm *,free-IMF}/L$. Each of the contours is closed, indicating that the black hole mass can be robustly constrained. 

From the $\chi^2$ distribution of the models with $M_{\rm *,const}/L$, we derived a black hole mass of $M_{\rm BH}= (8.4^{+1.3}_{-0.9})\times 10^7$ M$_{\odot}$ and a scaling factor S of $1.82 \pm 0.07$ at $3\sigma$ significance.  A comparison between the observed SINFONI and MUSE stellar kinematic maps and the model kinematics of our best-fitting model can be found in Figs. D.4. and D.5. of the Appendix. The counter rotation between the NSC and the host galaxy is very well recovered by our models. Also, the model velocity dispersion maps reproduce the central peak of the observations mostly well. For radii $> 0.5\arcsec$ in the SINFONI FoV, the model velocity dispersions are about 10\% higher than the observations but this is within the kinematic uncertainties. We notice that the features of the higher kinematic moments $h_3$ and $h_4$ cannot fully be reproduced by our models. As the residuals between the observed and modelled kinematics are small,  here the goodness of our model is strongly limited by the high observational uncertainties of the higher kinematic moments.

It is very common to combine the velocity and velocity dispersion maps to $V_{\rm rms} = \sqrt(V^2+\sigma^2)$ maps. We show the best-fit $V_{\rm rms}$ maps of FCC\,47 in Figure \ref{ff:bh_comparison} as they demonstrate the influence of the black hole mass on the model kinematics very
well. For the model with $M_{\rm *,const}/L$, the central peak, which spatially closely matches the NSC and reaches 151 km/s, is well recovered by our best-fit model. However, the $V_{\rm rms}$ outside of the NSC are slightly overestimated in our models.

For the stellar mass model obtained with $M_{\rm *,MW-IMF}/L$, we derive $M_{\rm BH}= (7.1^{+0.8}_{-1.1})\times 10^7 M_{\odot}$ and a scaling factor S of $0.83 \pm 0.04$ at $3\sigma$ significance. When also taking the spatial variation in IMF into account, we derive $M_{\rm BH}= (4.4^{+1.2}_{-2.1})\times 10^7 M_{\odot}$ and a scaling factor S of $0.86 \pm 0.04$ at $3\sigma$ significance. Also, for these best fits, we show the $V_{\rm rms}$ maps in Figure \ref{ff:bh_comparison}. As already seen in previous studies by \cite{Thater2019} and \cite{Thater2022}, the inclusion of a stellar $M_*/L$ gradient with fixed IMF affects the black hole mass, but only slightly. However, when the IMF is spatially varying within the centre of a galaxy (as is the case for FCC\,47), the stellar $M_*/L$ gradient can steepen, which leads to a non-negligible change in the estimated black hole mass. For FCC\,47, this change in black hole mass is almost 50\%. We note that the black hole mass derived with $M_{\rm *,free-IMF}/L$ has larger uncertainties compared to the models with $M_{\rm *,const}/L$ and $M_{\rm *,MW-IMF}/L$.

There is a straightforward explanation as to why we overestimate the black hole mass when assuming a constant $M_*/L$. Figure \ref{ff:ml} shows that there is a steep gradient in the observed stellar mass-to-light ratio: high values in the centre and lower values at larger radii. This means that there is in fact more stellar mass in the central region than is assumed when modelling a constant mass-to-light ratio for the full galaxy. In the models with constant mass-to-light ratio, more mass is assigned to the black hole to reproduce the observed kinematics, leading to an overestimation of the true black hole mass. This is more complicated due to the third contributor, dark matter, which is strongly degenerate with $M_*/L$ and black hole mass. However, for FCC 47, Figure \ref{ff:enclosed_mass} shows that (1) the enclosed mass profiles of the dark matter halos are very similar for the three mass models and (2) more importantly, the fraction of dark matter within $1\arcsec$ is negligible compared to the stellar and central black hole mass.

Also negligible is the systematic uncertainty due to the distance of the galaxy. The distance to FCC 47 was derived using surface-brightness-fluctuation measurements \citep{Blakeslee2009}, resulting in very low distance uncertainties. An uncertainty of 0.6 Mpc on the distance of 18.3 Mpc corresponds to only around 3\%.  Accounting for the systematic uncertainty due to distance leads to a black hole mass of $4.4^{+1.2}_{-2.1}({\rm stat}) \pm 0.14({\rm sys})\times 10^7 M_{\odot}$ for the measurement with varying IMF. As this systematic uncertainty is much smaller than the statistical uncertainty of our mass measurement, in the following we only quote the statistical uncertainty.

The black hole masses that we derived are only about 10\%-20\% of the NSC mass. The question arises as to whether the $V_{\rm rms}$ peak in Figure \ref{ff:bh_comparison} can also be recovered without the presence of a central black hole. We therefore also show the $V_{\rm rms}$ maps of models with $M_{\rm BH}= 0\,M_{\odot}$ and the best-fit mass-to-light ratio (fourth row in Figure \ref{ff:bh_comparison}). In this case, the $V_{\rm rms}$ would only reach 100 km/s for the model with constant $M_{\rm *,const}/L$ and 118 km/s for the model with $M_{\rm *,free-IMF}/L$ due to variation in stellar populations and IMF. Interestingly, the peak also slightly changes its orientation in the latter case. It is clear from these maps that the presence of a black hole is required to recover the $V_{\rm rms}$ in the centre of FCC\,47. If we were to mimic this increase in $V_{rms}$ simply by increasing the stellar $M_*/L$ in the centre of FCC\,47, this would require a value of about 4.5 M$_{\odot}/L_{\odot}$ assuming a constant mass-to-light ratio. We also show maps for this case in the bottom row of Figure \ref{ff:bh_comparison}. It becomes clear that such a high $M_*/L$ is then found everywhere in the centre and leads to a strong discrepancy with the SINFONI data. 

The construction of dynamical models also allowed us to estimate the mass of the NSC. Assuming that the stellar mass in the centre of FCC 47 is dominated by the NSC, we calculated the enclosed mass within the effective radius of the NSC ($R_{\rm eff,NSC} = 0.75\arcsec$). For the models with constant stellar $M_*/L,$ we obtained $M_{\rm NSC}= 3.2 \times 10^8 M_{\odot}$. This mass increased to $M_{\rm NSC}= 3.9 \times 10^8 M_{\odot}$ and $M_{\rm NSC}= 5.1 \times 10^8 M_{\odot}$ for $M_{\rm *,MW-IMF}/L$ and $M_{\rm *,free-IMF}/L$, respectively. Albeit slightly lower, our masses are consistent with the NSC mass derived by \cite{Fahrion2019} using stellar population models.

\begin{figure}
  \centering
      \includegraphics[width=0.48\textwidth]{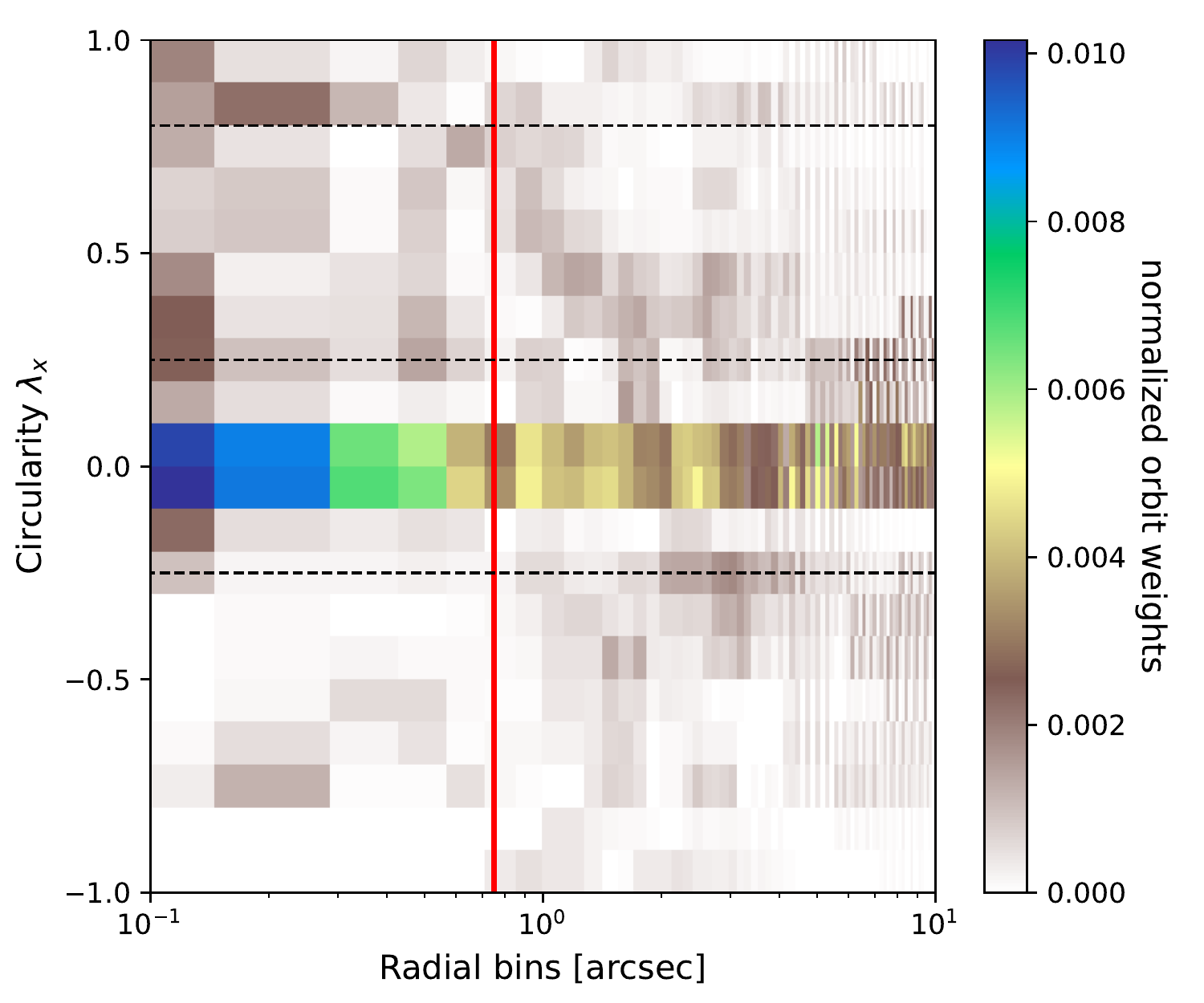}

      \includegraphics[width=0.48\textwidth]{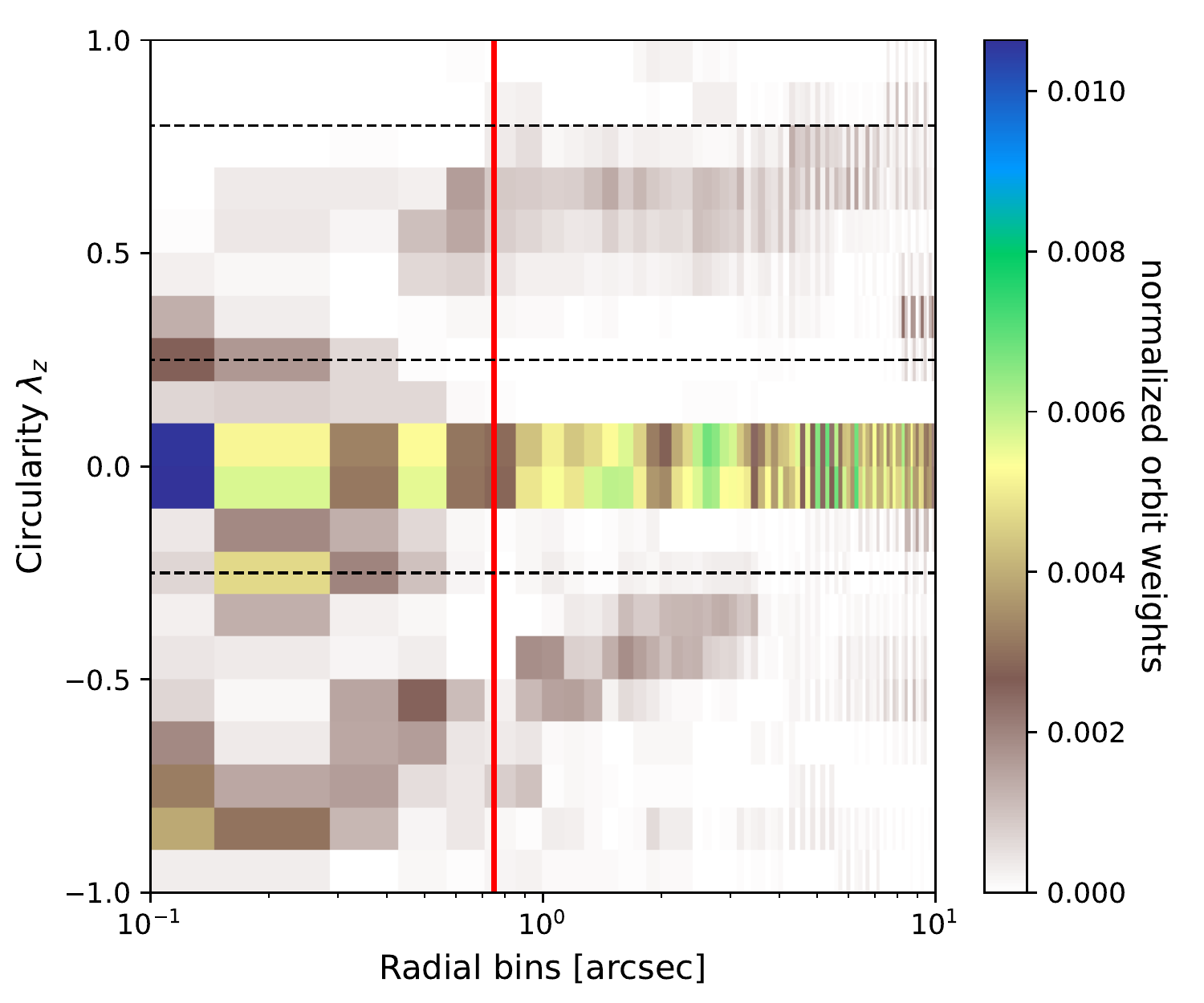}

      \caption{Averaged orbit circularity distribution of the ten best-fit models along the long (top) and short (bottom) galaxy axis. The colour indicates the weight of the orbits in phase-space. The dashed lines separate the distribution into cold ($\lambda_z>0.8$), warm ($0.25<\lambda_z<0.8$), hot ($|\lambda_z|<0.25$), and counter-rotating orbits ($\lambda_z<-0.25$). The red vertical line indicates the effective radius of the NSC.
      }
      \label{ff:lx}
\end{figure}

\subsection{The orbital distribution}

\cite{Lyubenova2019} showed that the NSC in FCC\,47 has very strong rotation ($|V|/\sigma \approx 0.5$) and high angular momentum ($\lambda_{\rm Re,NSC}=0.3$). This was confirmed by \cite{Fahrion2019} who showed with AO-assisted MUSE observations that the NSC rotates as a kinematically decoubled core in the galaxy with a rotation misalignment of 115$^{\circ}$ from the kinematic major axis of the galaxy. \cite{Fahrion2019} also showed that, based on the stellar orbit distribution, the rotation of the NSC can likely be disentangled from the rotation of the main galaxy. We refine their analysis by focussing on the very central scales that are dominated by the NSC and can be resolved with the SINFONI data (spatial resolution = $0\farcs14$). 

From the orbit-superposition Schwarzschild modelling of FCC 47, we constrain not only the gravitational potential but also the distribution of the orbital weights. The orbit distribution is often shown in the phase space of circularity and intrinsic radius of the orbit \citep{Zhu2018b}. The circularity, \mbox{$\lambda_{\rm z} = \overline{L_z} /(r \times \overline{V_c})$}, describes the orbit angular momentum $\overline{L_z}$ normalised by the angular momentum of a circular orbit with the same binding energy. This means that |$\lambda_{\rm z}$| = 1 represents highly rotating, dynamically cold short-axis tube orbits (circular orbits), while $\lambda_{\rm z} = 0$ represents mostly dynamically hot box or radial orbits. The circularity distribution has been used in the past to disentangle dynamical cold, warm, and hot components and to gain insight into the accretion history of nearby galaxies \citep[e.g.][]{Zhu2018b,Zhu2020,Zhu2022,Fahrion2019,Poci2019}.

Figure \ref{ff:lx} shows the circularity distribution along the long and short axes. We derived the distribution from our ten best-fit dynamical models using the mass model with $M_{\rm *,const}/L$. The circularity distributions of the other two mass models have the same general features. This is because differences in the circularity distribution are mostly driven by the stellar kinematics. Compared to Figure 13 and 14 of \cite{Fahrion2019}, which show a more global orbital distribution, we here zoom into the central $0.1''-10'' = 9-900$ pc. All the way to the very centre of FCC\,47, the orbital distribution of FCC\,47 is dominated by hot orbits ($\lambda_{\rm z} \approx 0$). In ETGs like FCC\,47, hot orbits typically dominate the orbit distribution \cite[e.g.][]{Santucci2022}. In FCC\,47, we find two additional regions with a large fraction of orbits in the circularity distribution. 

Along the long axis, we see a distinct set of warm and cold orbits within the central $0.5''$ (Figure \ref{ff:lx}, top). As there are almost no counter-rotating orbits in this region, this results in prolate net rotation along the long axis within $0.5''$. Along the short axis, we see a large fraction of counter-rotating orbits out to $1''$ (Figure \ref{ff:lx}, bottom). 
\cite{Fahrion2019} also found these two distinct regions; however, due to the limitation in spatial resolution of the MUSE data, these authors were only able to limit the rotation along the long axis to the central $4''$. Here, we can see that, in reality, the rotation originates from the central $0.4''$, where the NSC dominates the kinematics. In our follow-up project, we plan to use the information of the circularity distribution in combination with orbit colouring to disentangle the NSC from the main galaxy. 

\section{Discussion}
\label{s:discussion}

\subsection{Systematic effects of including the (non-)variable IMF}
Using triaxial DYNAMITE Schwarzschild modelling, we measured the SMBH mass in the nucleated galaxy FCC\,47. We obtained $M_{\rm BH} = (8.4^{+1.3}_{-0.9})\times 10^7 M_{\odot}$ using a mass model with constant $M_*/L$,  and  $M_{\rm BH} = (4.4^{+1.2}_{-2.1})\times 10^7 M_{\odot}$ using a mass model with a varying $M_*/L$ due to spatially varying age, metallicity and IMF. Recently, there has been an increasing number of studies that account for a varying $M_*/L$ with fixed IMF in their dynamical modelling of stellar kinematics \citep{McConnell2013b,Nguyen2017a,Nguyen2018,Nguyen2019a,Thater2019,Thater2022,Poci2021} and molecular gas kinematics \citep[e.g.][]{Davis2017,North2019,Nguyen2020,Nguyen2021,Nguyen2021b}. When also modelling the central black hole, these studies reported a decrease in black hole mass that was usually about 10\%-15\% and therefore insignificant with respect to the mass measurement uncertainties. 

Here, we present a dynamical measurement of black hole mass that takes into account a varying $M_*/L$ due to spatial variation in age, metallicity, and, for the first time, IMF. This extension was crucial for FCC\,47 as we noticed significant changes of stellar populations and IMF slope in the centre of FCC 47, where the NSC dominates. Consequently, considering IMF variation in the dynamical modelling, the black hole mass decreased by almost~50\%. 

This change in the black hole mass is a consequence of the impact of a variable IMF on the stellar mass-to-light ratio, which is two-fold. First, an IMF with a relative excess of low-mass stars (bottom-heavy), as found in the central regions of FCC\,47, would lead to a higher mass-to-light ratio than expected for a Milky Way-like IMF. Complementarily, a relatively old population with an IMF biased towards massive stars (top-heavy) would also have a higher mass-to-light ratio but one that is driven by the presence of dark remnants (i.e. stellar black holes and neutron stars). While the effect of the remnants is heavily dependent on the assumed IMF parametrisation and cannot be constrained from stellar population analyses, the impact of a variable number of low-mass stars can be approximated by fitting the optical absorption spectra of FCC\,47, as shown in Figure 2. Due to the central spatial variation in IMF, the central $M_*/L$ increases from about 2.7 M$_{\odot}/$L$_{\rm \odot}$ (assuming a constant MW-like IMF) to about 3.3 M$_{\odot}/$L$_{\rm \odot}$ (allowing for spatial variation in the IMF).
If we were to fail to take into account the IMF in our dynamical modelling, the central mass-to-light ratio would be lower, and therefore less mass would be assigned to the stars while more mass would be assigned to the SMBH.

The inclusion of IMF variation may have an impact on other dynamical black hole mass estimates as well. 
Adaptive-optics-assisted MUSE observations make it possible to obtain detailed maps of the IMF.  \cite{Martin-Navarro2015a}, \cite{Labarbera2016}, \cite{Sarzi2018}, \cite{Martin-Navarro2019},  \cite{Martin-Navarro2021}, and  \cite{Barbosa2021} show strong variations in IMF for massive galaxies like M87 as well. Additional dynamical studies are needed in the future in order to understand the implications of our findings on black hole estimates and black hole scaling relations. 

\begin{figure}
  \centering
      \includegraphics[width=0.49\textwidth]{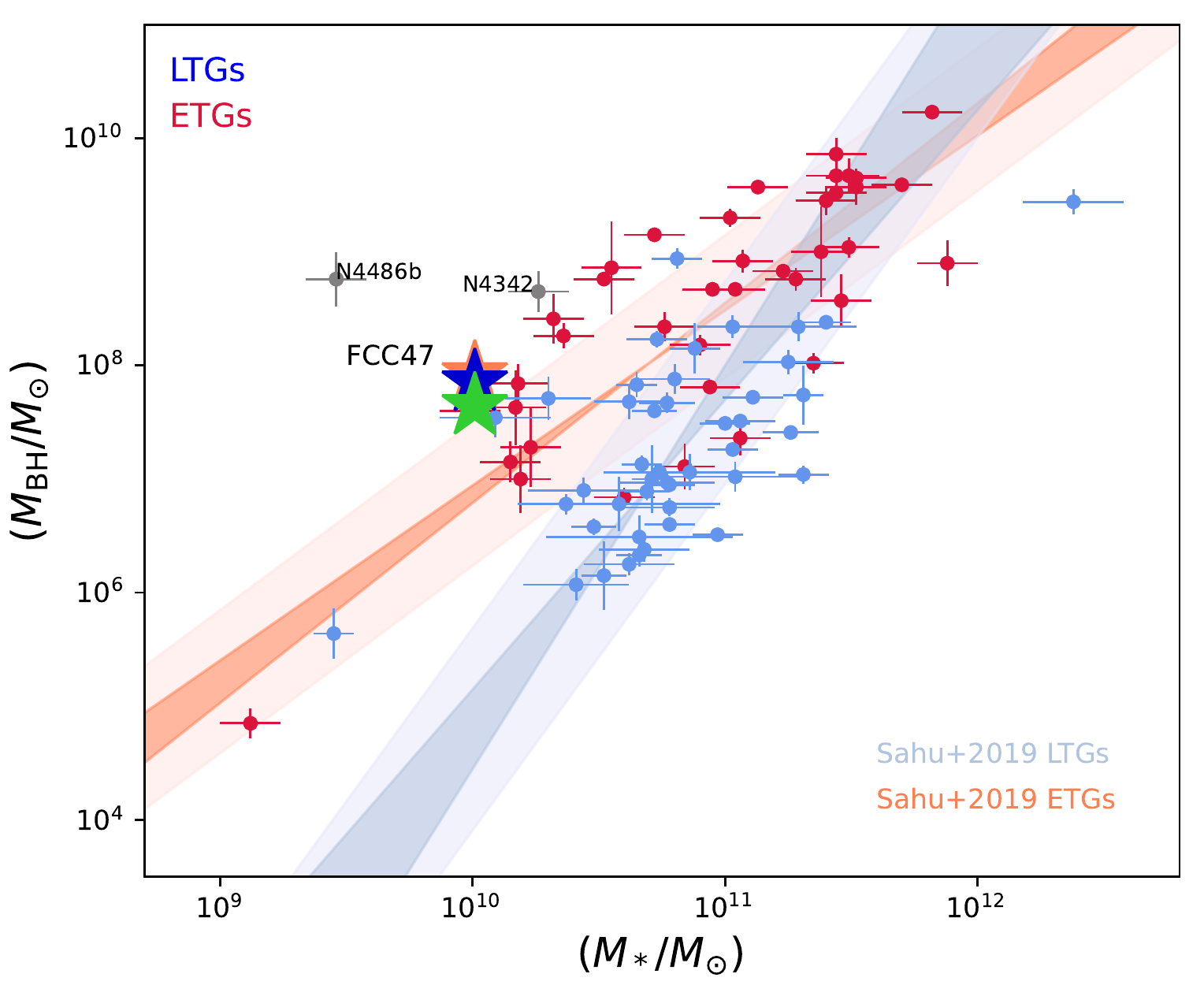}

      \caption{Relation between central black hole mass and stellar mass of the host galaxy. The plot is adapted from \cite{Sahu2019}. The data points were taken from \cite{Sahu2019} for ETGs and \cite{davis2018b} for LTGs. NGC 4486b and NGC4342 are tidally stripped galaxies (with significantly decreased host galaxy mass) and are therefore masked in grey. The scaling relations for ETGs and LTGs are visualised as shaded areas. For the ETG FCC 47, the colours indicate the measurement method of this study: orange indicates constant $M_*/L$, blue is for varying $M_*/L$ with fixed MW-like IMF, and green is for varying $M_*/L$ with free IMF.
      }
      \label{ff:bh_gal}
\end{figure}

\begin{figure}
  \centering
      \includegraphics[width=0.49\textwidth]{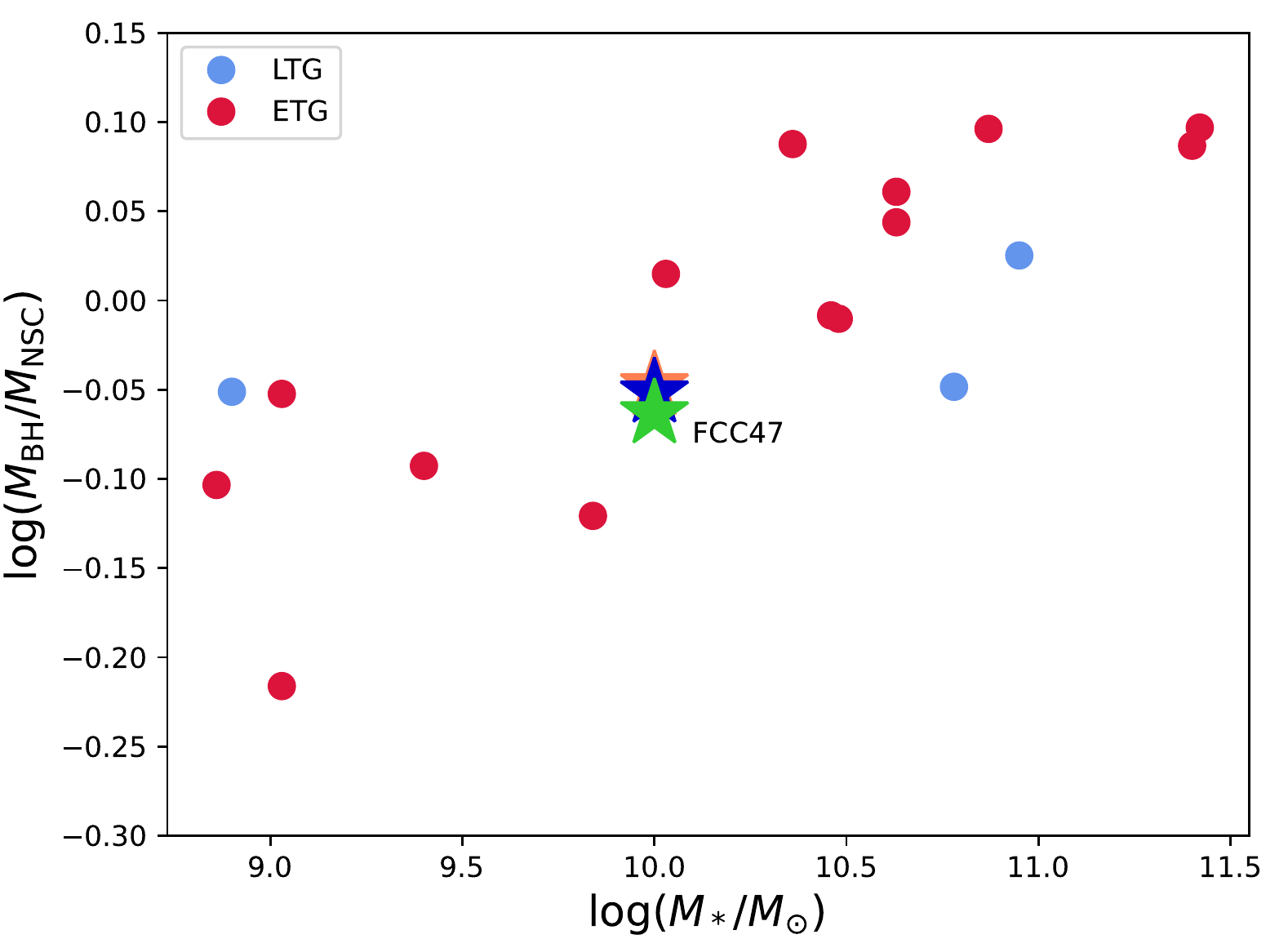}

      \caption{Ratio of the central black hole mass and the NSC mass plotted against their host galaxy stellar mass. The figure is adapted from \cite{Neumayer2020}. We only plot galaxies that have a robust measurement in both black hole and NSC mass. The colours indicate the galaxy type: red for ETGs and blue for LTGs. For the ETG FCC 47, the colours have the same significations as in Fig. 11.
      }
      \label{ff:bh_nsc}
\end{figure}

\subsection{Is the SMBH in FCC\,47 as over-massive as the NSC?}

Our measurement in FCC\,47 can be compared to other galaxies via various NSC and SMBH scaling relations. Figure \ref{ff:bh_gal} shows the black hole compilation by \cite{Sahu2019}. This authors divide their sample into early- and late-type galaxies (LTGs) and find two different scaling relations. As FCC\,47 is an S0 galaxy, we focus here on the scaling relation of ETGs (red).
It is immediately clear that in the galaxy-mass range of FCC\,47, this plot is only marginally populated at present. Therefore, the red scaling relation is also not well anchored on black hole mass measurements in this regime.  For FCC\,47 to follow this relation, it would need a black hole mass of about $M_{\rm BH}= 5\times 10^6 M_{\odot}$, 13 times less massive. This means that, compared to this scaling relation, our black hole is strongly over-massive. Our mass measurement with varying IMF only slightly pushes the galaxy closer to the scaling relation.

However, when only focusing on the measurements at around $M_{\rm *,gal}\sim10^{10}$ M$_{\odot}$, it is clear that the red scaling relation is not a good description of these measurements. About half of the ETGs in this mass regime are significantly above the red relation; for example, NGC 307 \citep{Erwin2018}, NGC 1374 \citep{Saglia2016}, NGC 5845 \citep{Schulze2011a}, NGC 4434, and NGC 4339 \citep{Krajnovic2018}. It is therefore possible that the scaling relation for ETGs has a shallower slope than currently debated in the literature. Further insight will be provided by increasingly precise black hole mass measurements for low-mass ($M_{\rm *,gal} < 10^{10}$ M$_{\odot}$) ETGs. The same trend is seen when adding our galaxy to the black hole--sigma scaling relation.
We therefore conclude that the black hole in FCC\,47 is probably not over-massive with respect to its host galaxy mass. 

We also calculated the mass ratio between the SMBH and NSC in FCC\,47. The measurement with $M_{\rm *,const}/L$ provides $\log(M_{\rm BH}/M_{\rm NSC}) = -1.0$ and with IMF variation, $\log(M_{\rm BH}/M_{\rm NSC}) = -1.2$. Figure \ref{ff:bh_nsc} shows how FCC\,47 compares to the literature. Our galaxy aligns well with the other measurements but is NSC-dominated. A scenario where NSC has outgrown the massive black hole of the galaxy is imaginable. \cite{Fahrion2019} and \cite{Fahrion2022} investigated the star formation history of the NSC in FCC\,47, finding it to be dominated by an old, metal-rich population. Based on this, the authors concluded that the NSC of FCC\,47 formed  in situ early-on through efficient star formation following accretion of gas. This gas could also have fed the black hole, but this feeding was shut off at some point before it could grow over-massive. The lack of stellar populations younger than $\sim$ 10 Gyr supports this scenario, in which the star formation in the NSC ---and the black hole growth--- was stopped a long time ago.

\section{Conclusion}
In this study we present our black hole mass estimate in the nucleated ETG FCC\,47. We combined HST/ACS F850LP band data from the ACSFCS survey with AO-assisted K-band SINFONI and AO-assisted optical MUSE IFU observations. The NSC in the centre of FCC\,47 clearly reveals itself as a kinematically decoubled core (r $\approx$ 0.7") in both IFU observations. The velocity dispersion is peaked in the very centre, indicating the presence of a black hole. We performed triaxial Schwarzschild orbit-superposition modelling in order to constrain the intrinsic shape parameters, the dark matter distribution (parametrised as NFW), the stellar mass, and the mass of the central black hole of FCC\,47. Detailed knowledge about the stellar mass is crucial for the accuracy of the black hole mass. Traditionally, the stellar mass is calculated with a constant $M_*/L$ in dynamical models. This is a poor assumption for FCC\,47 as we show that the NSC residing in the centre of FCC\,47 has a very different stellar $M_*/L$ from the host galaxy. We improved the traditional approach by implementing a varying $M_*/L$ due to variation in age, metallicity, and, for the first time, IMF. Based on our modelling, we find the following:
\begin{itemize}
    \item We can clearly exclude Schwarzschild models without a central black hole. Consequently, our models predict that a massive black hole resides within the NSC of FCC 47.
    \item When assuming a constant stellar $M_*/L$, we derive a black hole mass $M_{\rm BH} = (8.4^{+1.3}_{-0.9})\times 10^7 M_{\odot}$ and a dynamical $M_*/L$ of 1.82 M$_{\odot}$/L$_{\odot}$.
    \item The black hole mass decreases by 15\% when a stellar-population-based spatially varying $M_*/L$ with fixed MW-IMF is assumed instead. We measure 
    $M_{\rm BH} = (7.1^{+0.8}_{-1.1})\times 10^7 M_{\odot}$. Furthermore, we derive a mass scaling factor (the ratio between dynamical and stellar-population-based $M_*/L$) of 0.83, which indicates that  our stellar population analysis probably predicts an excessively high $M_*/L$. The difference in black hole mass when assuming a varying $M_*/L$ with fixed IMF is within the measurement uncertainties.
    \item The black hole mass decreases by almost 50\% when a stellar-population-based spatially varying $M_*/L$ with spatially varying IMF is taken into account in the dynamical models. We measure 
    $M_{\rm BH} = (4.4^{+1.2}_{-2.1})\times 10^7 M_{\odot}$. We note that the uncertainties are larger for this mass measurement, which is probably due to larger uncertainties in the mass model. This time, we derive a mass scaling factor of 0.86. A rule of thumb suggest a general uncertainty of a factor of two for black hole mass measurements. When taking IMF variations into account, the change in black hole mass is of the same order as the measurement uncertainties. More measurements like this will be important in the future to understand whether ignoring the IMF variations might explain some of the scatter in the black hole scaling relations.

\end{itemize}

As for FCC 47, the NSC has been well studied in previous literature. We add to those measurements a robust black hole mass measurement. Comparison to the literature suggests that while the NSC is very over-massive compared to the host-galaxy, the central black hole is probably not. Our findings suggest that the NSC formed very early through very efficient star formation after a period of  gas accretion. However, this gas was not sufficient enough to grow an over-massive black hole. In the future, we will combine our dynamical models with age and metallicity information to better understand how the centre of FCC\,47 was built up.

\section*{Acknowledgements}
We thank the anonymous referee for helpful comments and suggestions that improved this manuscript. We also thank Mark den Brok, Adriano Poci, Giulia Santucci and Francesca Pinna for valuable discussions, and acknowledge the great support from Alice Zocchi and the Vienna Dynamics Team. This work is based on observations collected at the European Organization for Astronomical Research in the Southern Hemisphere under ESO programmes 92.B-0892 and 60.A-9192. S.T. and M.L. acknowledge funding from the ESO Science Support Discretionary Fund. KF acknowledges support through the ESA research fellowship programme. This research was supported by the European Union's Horizon 2020 research and innovation programme under grant agreement NO 724857 (Consolidator Grand ArcheoDyn). DDN is grateful to the LABEX Lyon Institute of Origins (ANR-10-LABX-0066) Lyon for its financial support within the program ``Investissements d'Avenir'' of the French government operated by the National Research Agency (ANR). The computational results presented have been achieved in part using the Vienna Scientific Cluster (VSC).
\\ \\
Software: DYNAMITE \citep{Jethwa2020,Thater2022b}, pPXF \citep{Cappellari2004,Cappellari2017}, vorbin \citep{Cappellari2003}, MgeFit \citep{Cappellari2002}, plotbin,  astropy \citep{Astropy2022}, scipy \citep{Virtanen2020}

\bibliography{papers} 
\bibliographystyle{aa}

\begin{appendix}
\section{Determination of the SINFONI and HST spatial resolution}
\subsection{SINFONI PSF}
\begin{figure}
  \centering
        \includegraphics[width=0.50\textwidth]{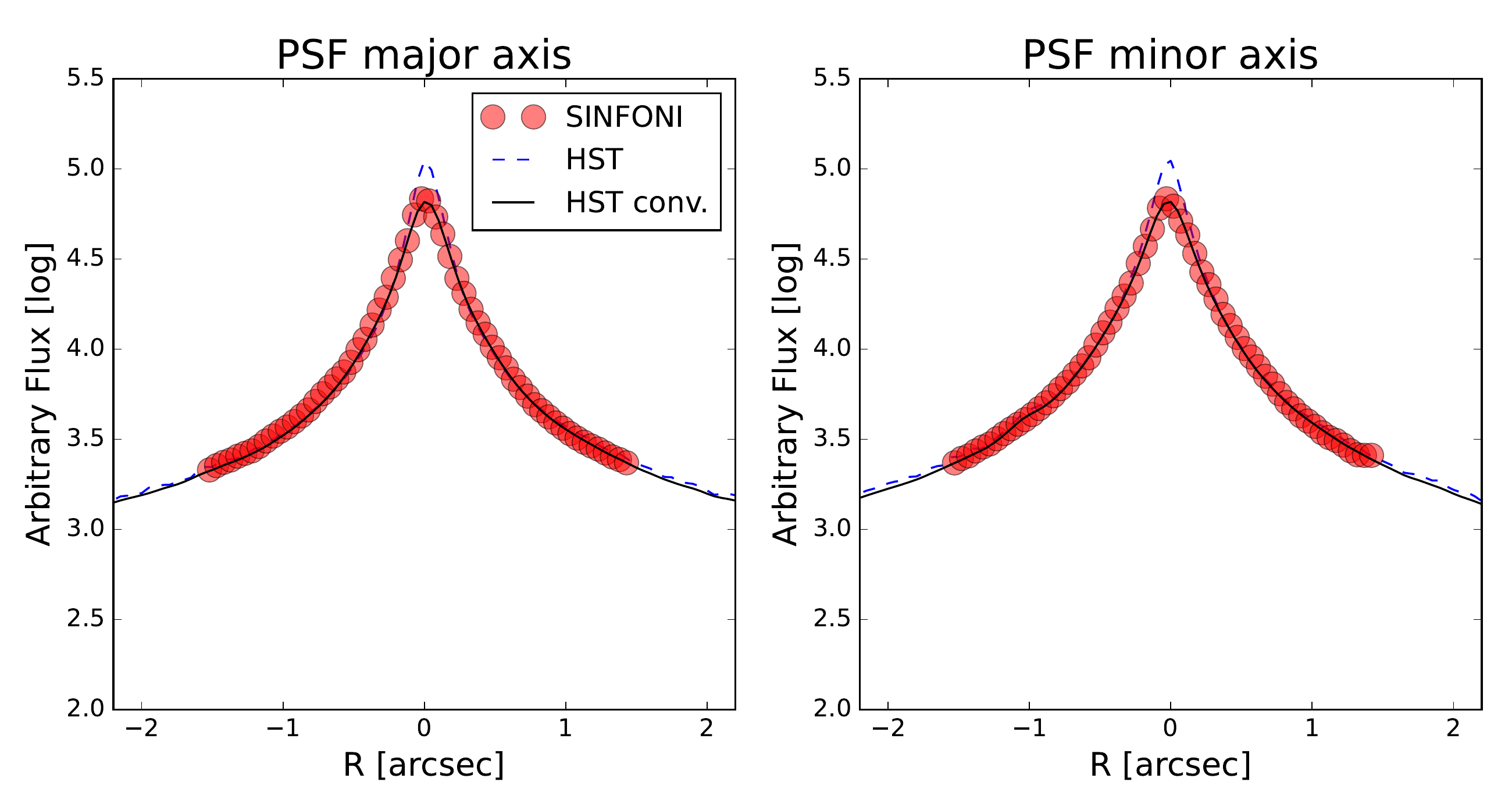}

      \caption{Spatial resolution of our SINFONI observations derived by comparison
of the SINFONI white-light image with the HST/ACS F850 image.}
      \label{ff:psf}
\end{figure}

For estimating the spatial resolution of the SINFONI observations, we used the same approach as in \cite{McDermid2006}, \cite{Thater2017,Thater2020} and \cite{Lyubenova2019}. The idea is to use a high-resolution image of the galaxy at similar wavelength and degrade it via PSF convolution until it matches the white-light image of the IFU observations. We did not have an image in the near-infrared and therefore used the reddest HST image, F850LP. We then convolved the full HST image with a toy PSF parametrised as the sum of two concentric Gaussians (see Figure \ref{ff:psf}). The convolved HST image best matches the reconstructed SINFONI observations with a narrow Gaussian of 0\farcs14 (FWHM) and a broad Gaussian of 0\farcs5 (FWHM). The narrow Gaussians contributes 67.5 \%. These values are consistent with the PSF measurement by \cite{Lyubenova2019}, who parametrised the PSF as a single Gaussian and obtained about 0\farcs2.

\subsection{HST PSF}
We generated an image of the HST/ACS PSF using the Tiny Tim HST PSF modelling tool \citep{Krist2001} taking into account the imaging filter, the central position of the galaxy on the CCD chip, and assuming the spectrum of a K giant star. The PSF was modelled by a sum of concentric circular Gaussians using the MGE method \citet{Cappellari2002}. The relative weight of the Gaussians was normalised such that the sum of the weights equals one. The MGE parameters of each Gaussian is provided in Table \ref{tt:hst_psf}. When modelling the Tiny Tim PSF with a single circular Gaussian, we obtained ${\rm FWHM = 0.09''}$. In Section 3, we used the PSF parametrisation given in Table \ref{tt:hst_psf}.
\begin{table}
\caption{MGE parameters of the HST/ACS/F850LP circular PSF fit}
\centering
\begin{tabular}{ccc}
\hline\hline
k & $G_k$  &  $\sigma_k^*$ \\
 &   &  (arcsec) \\
\hline
1 & 0.298 & 0.0268 \\
2 & 0.576 & 0.0678 \\
3 & 0.035 & 0.1335\\
4 & 0.091 & 0.3043\\
\hline
\end{tabular}
\label{tt:hst_psf}
\end{table}

\section{MGE parametrisation}
Tables of the Multi-Gaussian Expansions that are discussed in Section 3.
\begin{table}
\caption{HST/ACS F850LP 
luminosity MGE model.}
\centering
\begin{tabular}{lcccrr}
\hline\hline
j &   $\log\,(M_{j}^{\rm const}$) & $\log\,(I_j$) & $\sigma_j$ & q$_j$ & $\Psi_j$ \\ 
 & (M$_{\sun}$) & (L$_{\sun}$ pc$^{-2}$) & (arcsec) & & (deg)  \\ 
(1)  & (2) &  (3) & (4) & (5) & (6)  \\
\hline
1  & 6.298 & 4.603 & 0.024 & 0.99 & 0.0 \\
2  & 6.037 & 4.163 & 0.038 & 0.57 & 0.0 \\
3  & 7.74 & 4.952 & 0.088 & 0.87 & 90.0 \\
4  & 7.878 & 4.601 & 0.17 & 0.72 & 79.76 \\
5  & 7.919 & 3.793 & 0.467 & 0.68 & -90.0 \\
6  & 7.992 & 3.476 & 0.604 & 1.0 & 22.86 \\
7  & 8.333 & 3.219 & 1.278 & 0.89 & 69.15 \\
8  & 8.615 & 2.925 & 2.429 & 0.92 & -30.59 \\
9  & 7.77 & 1.708 & 3.739 & 0.92 & -6.69 \\
10  & 8.804 & 2.504 & 4.993 & 0.89 & -30.59 \\
11  & 9.066 & 2.215 & 9.296 & 0.91 & -34.53 \\
12  & 7.613 & 0.247 & 22.958 & 0.49 & -0.01 \\
13  & 9.373 & 1.845 & 23.982 & 0.65 & -35.27 \\
14  & 9.696 & 1.312 & 56.037 & 0.86 & -31.75 \\

\hline
\\
\end{tabular}
\\
\tablefoot{Column 1: Index of the Gaussian component. Column 2: Mass of Gaussian component derived by multiplying the luminosity of each Gaussian with the constant dynamical $M_*/L$ = 1.82\,M$_{\odot}$/L$_{\odot}$ from the Schwarzschild modelling. Column 3: Surface brightness. Column 4: Gaussian width along the  axis. Column 5: Axial ratio for each Gaussian component. Column 6: Position angle for each Gaussian component. The stellar mass of the galaxy based on this MGE is $1.02 \times 10^{10}$ M$_{\odot}$. The effective radius derived from this MGE is 29.3 arcsec.  }
\label{tt:mge}
\end{table}

\begin{table}
\caption{HST/ACS F850LP 
mass MGE model assuming a varying stellar $M_*/L$ and fixed MW-like IMF.}
\centering
\begin{tabular}{lcccrr}
\hline\hline
j &   $\log\,(M_{j}^{\rm const}$) & $\log\,(\Sigma_j$) & $\sigma_j$ & q$_j$ & $\Psi_j$ \\ 
 & (M$_{\sun}$) & (M$_{\sun}$ pc$^{-2}$) & (arcsec) & & (deg)  \\ 
(1)  & (2) &  (3) & (4) & (5) & (6)  \\
\hline
1  & 6.758 & 5.081 & 0.034 & 1.0 & 90.00 \\
2  & 7.805 & 5.364 & 0.082 & 1.0 & 90.00 \\
3  & 7.953 & 5.067 & 0.171 & 0.64 & 90.00 \\
4  & 7.973 & 4.283 & 0.409 & 0.71 & 81.61 \\
5  & 7.721 & 3.607 & 0.563 & 1.0 & 14.89 \\
6  & 8.159 & 3.827 & 0.796 & 0.82 & -87.19 \\
7  & 8.427 & 3.521 & 1.402 & 1.0 & -18.87 \\
8  & 8.507 & 3.107 & 2.475 & 1.0 & -0.01 \\
9  & 8.69 & 2.799 & 4.395 & 0.98 & -4.92 \\
10  & 8.2 & 2.585 & 4.525 & 0.49 & -33.96 \\
11  & 8.997 & 2.578 & 8.4 & 0.91 & -30.24 \\
12  & 9.302 & 2.223 & 20.185 & 0.72 & -36.77 \\
13  & 9.532 & 1.431 & 56.046 & 0.98 & -69.4 \\
14  & 9.161 & 1.361 & 56.046 & 0.49 & -28.08 \\

\hline
\\
\end{tabular}
\\
\tablefoot{Column 1: Index of the Gaussian component. Column 2: Mass of Gaussian component derived with the constant dynamical $M_*/L$ = 0.83\,M$_{\odot}$/L$_{\odot}$ from the Schwarzschild modelling. Column 3: Surface mass density. Column 4: Gaussian width along the  axis. Column 5: Axial ratio for each Gaussian component. Column 6: Position angle for each Gaussian component. The stellar mass of the galaxy based on this MGE is $1.0 \times 10^{10}$ M$_{\odot}$.  }
\label{tt:mge2}
\end{table}

\begin{table}
\caption{HST/ACS F850LP 
mass MGE model assuming a varying stellar $M_*/L$ and varying IMF.}
\centering
\begin{tabular}{lcccrr}
\hline\hline
j &   $\log\,(M_{j}^{\rm const}$) & $\log\,(\Sigma_j$) & $\sigma_j$ & q$_j$ & $\Psi_j$ \\
 & (M$_{\sun}$) & (M$_{\sun}$ pc$^{-2}$) & (arcsec) & & (deg)  \\ 
(1)  & (2) &  (3) & (4) & (5) & (6)  \\
\hline
1  & 7.013 & 5.262 & 0.037 & 1.0 & 90.00 \\
2  & 7.619 & 5.204 & 0.08 & 1.0 & 90.00 \\
3  & 8.033 & 5.339 & 0.126 & 0.77 & 88.64 \\
4  & 7.862 & 4.657 & 0.262 & 0.57 & 74.5 \\
5  & 8.335 & 4.282 & 0.568 & 0.86 & 71.65 \\
6  & 8.467 & 3.659 & 1.304 & 0.93 & 74.26 \\
7  & 8.402 & 3.126 & 2.209 & 0.96 & -27.53 \\
8  & 8.197 & 2.684 & 4.053 & 0.49 & -34.0 \\
9  & 8.672 & 2.864 & 4.221 & 0.89 & 47.89 \\
10  & 8.754 & 2.543 & 7.219 & 0.77 & -38.1 \\
11  & 8.838 & 2.278 & 9.532 & 0.99 & -0.62 \\
12  & 9.31 & 2.208 & 21.6 & 0.67 & -36.66 \\
13  & 9.133 & 1.338 & 56.037 & 0.49 & -30.62 \\
14  & 9.563 & 1.458 & 56.037 & 1.0 & -63.69 \\

\hline
\\
\end{tabular}
\\
\tablefoot{Column 1: Index of the Gaussian component. Column 2: Mass of Gaussian component derived with the constant dynamical $M_*/L$ = 0.86\,M$_{\odot}$/L$_{\odot}$ from the Schwarzschild modelling. Column 3: Surface mass density. Column 4: Gaussian width along the  axis. Column 5: Axial ratio for each Gaussian component. Column 6: Position angle for each Gaussian component. The stellar mass of the galaxy based on this MGE is $1.0 \times 10^{10}$ M$_{\odot}$.}
\label{tt:mge3}
\end{table}

\section{Extrapolated $M_*/L$ maps in the i-band}
\begin{figure*}
  \centering
        \includegraphics[width=0.50\textwidth]{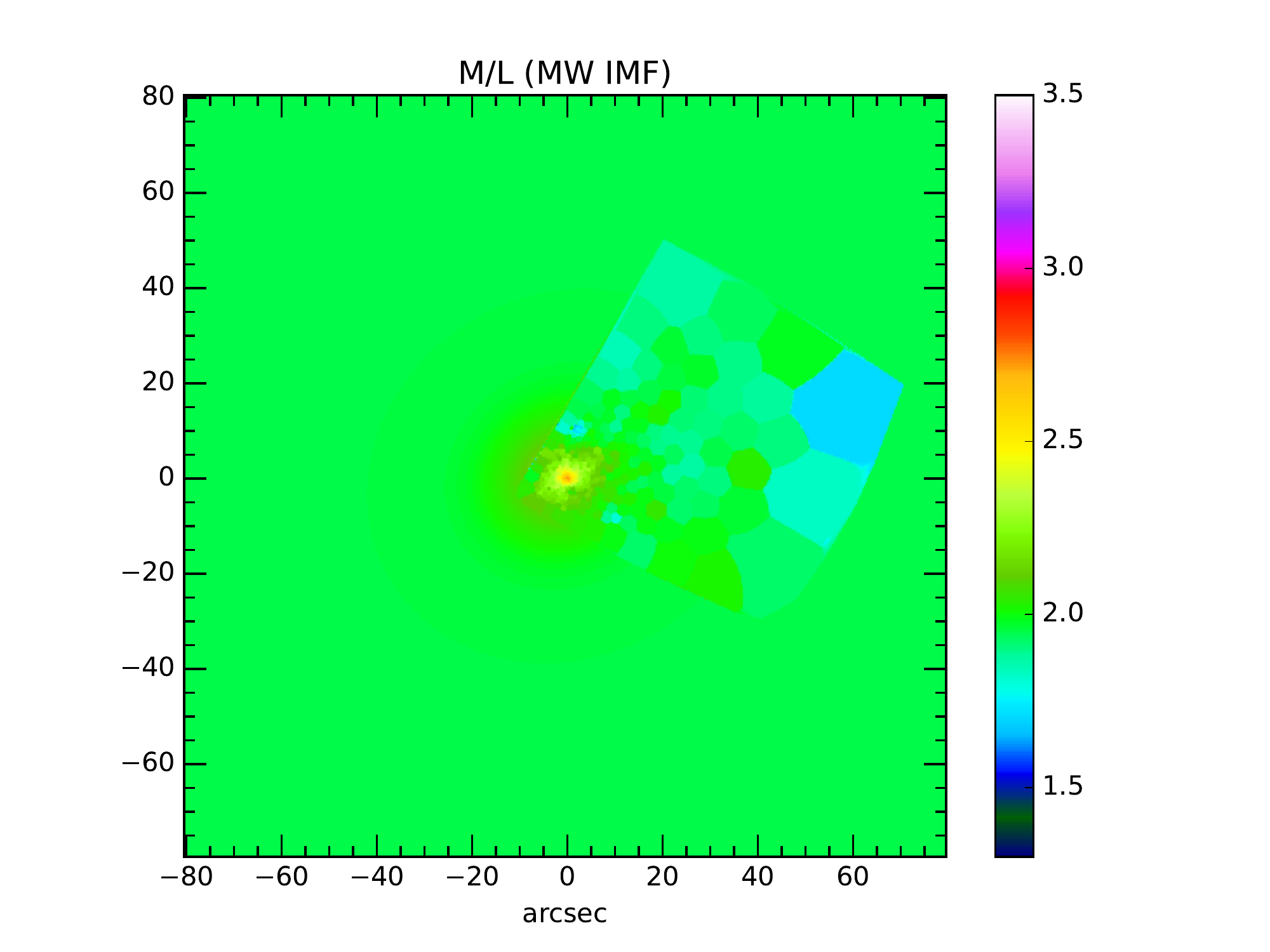}
        \includegraphics[width=0.48\textwidth]{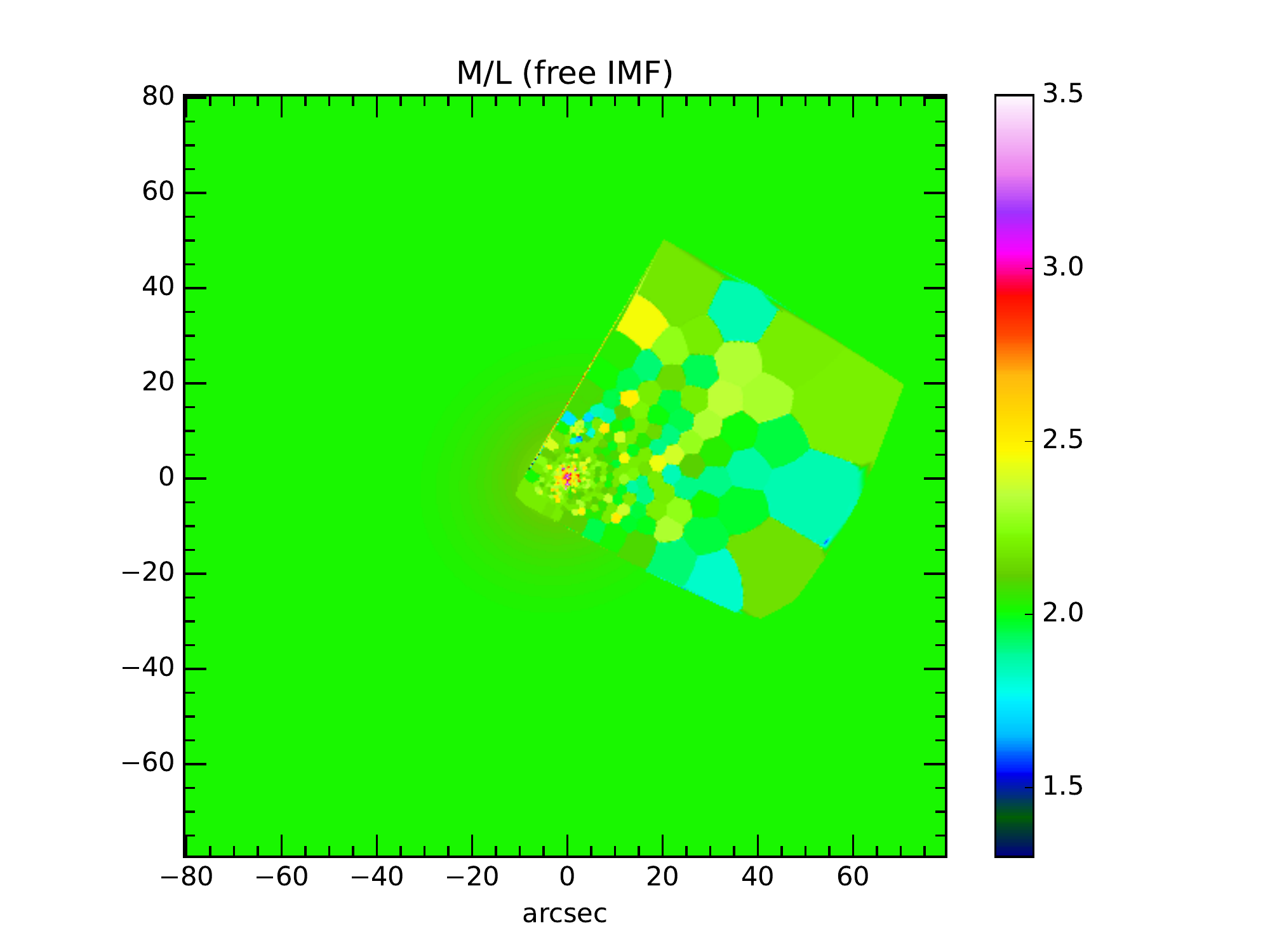}

      \caption{Extrapolated $M_*/L$ map for the full ACS FOV. The original $M_*/L$ maps were derived from MUSE single stellar populations using a constant MW-like IMF (left) and a varying IMF. 
      }
      
\end{figure*}

\section{Best-fit DYNAMITE models}

  \begin{table*}
  \label{tt:results}
\caption{Results of the iterative parameter search for Schwarzschild models with different stellar mass models.}
\centering
\begin{tabular}{lcccccccc}
\hline\hline
Mass model  & $\log (M_{\rm BH}$/M$_{\odot}$)  &  $S$ & $p$ & $q$ & $u$ & $\log f_{\rm DM}$ & $\log c$ & $\chi^2$/d.o.f.\\
 (1) & (2) & (3) & (4) & (5) & (6) & (7)  & (8)    \\
\hline
$M_{\rm *,const}/L$ & $7.90\pm 0.1$ & *1.80$\pm 0.20$ & 0.59$\pm 0.06$ & 0.23$\pm 0.08$ & 0.93$\pm 0.03$ & 1.25$\pm 0.30$ & 1.19$\pm 0.20$ & 2.18 \\
$M_{\rm *,MW-IMF}/L$ & $7.85 \pm 0.1$ & 0.83$\pm 0.04$ & $0.59 \pm 0.05$  & $0.18 \pm 0.08$ & $0.93\pm 0.03$ & $1.75\pm 0.30$ & $0.95\pm 0.20$ & 2.17\\
$M_{\rm *,free-IMF}/L$ & $7.60 \pm 0.2$ & 0.89$\pm 0.04$ & $0.55\pm 0.10$ &  $0.20\pm 0.05$ &  $0.95\pm 0.05$ & $1.78 \pm 0.90$ & $0.80\pm 0.30$ & 2.18\\
\hline
\end{tabular}
\\
\tablefoot{This table summarises the results of Section 4.2. Column 1: Details of the stellar mass model used for the dynamical modelling. Columns 2 and 3: Parameters of the best-fitting models derived from the
regular grid search: black hole mass MBH and scaling factor S. Columns 4 to 8: Parameters of the best-fitting models to constrain the global gravitational potential: p (intrinsic medium-to-major axis ratio), q (intrinsic minor-to-major axis ratio), u (ratio between projected and intrinsic major axis), the dark matter fraction at $r_{200}$, $f_{\rm DM}$, and the dark matter concentration c. 
Column 9: The $\chi^2$ over the degrees of freedom. 
*For the model with constant mass-to-light ratio the mass scaling factor S corresponds to the derived stellar $M_*/L_{\rm F850}$ in M$_{\odot}/$L$_{\rm \odot}$.}
\end{table*}

\begin{figure*}
  \centering
      \includegraphics[width=1\textwidth]{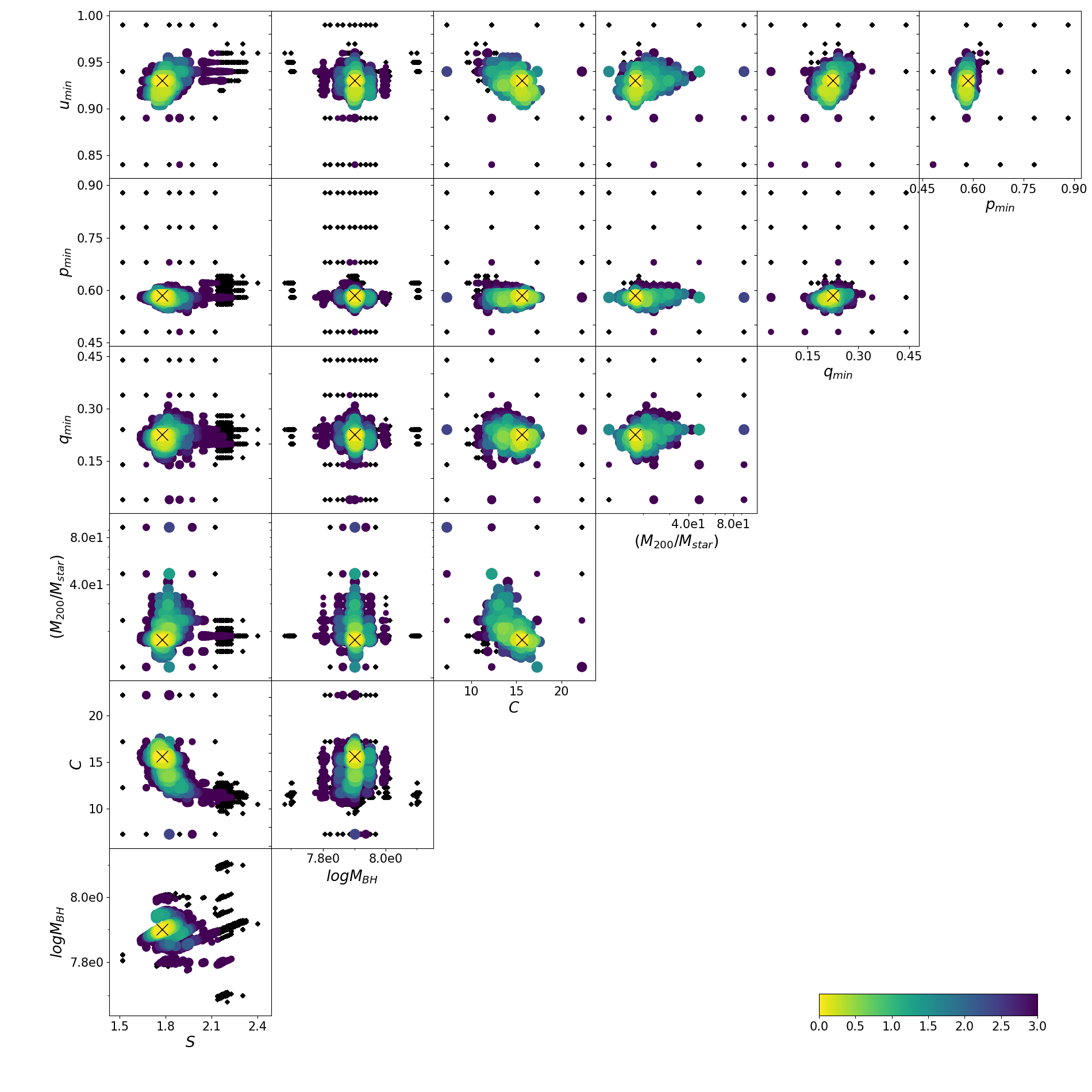}

      \caption{$\chi^2$ distribution of our dynamical models to constrain intrinsic shape and dark matter parameters for the mass model with constant $M_*/L$. The seven hyper-parameters are the mass scaling factor S (in solar units), the black hole mass in solar units, the dark matter concentration c, and dark matter halo mass in units of the stellar mass $M_{200}/M_{star}$, the intrinsic minor-to-major axis ratio q, the intrinsic medium-to-major axis ratio p and the ratio between projected and intrinsic major axis u. Each point is one model that is colour-coded according to its $\chi^2$ value as shown in the colour bar, where ($\chi^2$-$\chi_{\rm min}^2)/\sqrt{2n_{\rm GH}\times N_{\rm kin}} < 1$ indicates the models within 1$\sigma$ confidence interval. Here, $n_{\rm GH}$ corresponds to the number of fitted Gauss-Hermite moments (4) and $N_{\rm kin}$ is the number of kinematic bins. The black cross marks the best-fitting model.
      }
      \label{ff:chi2_global}
\end{figure*}

\begin{figure*}
  \centering
      \includegraphics[width=1\textwidth]{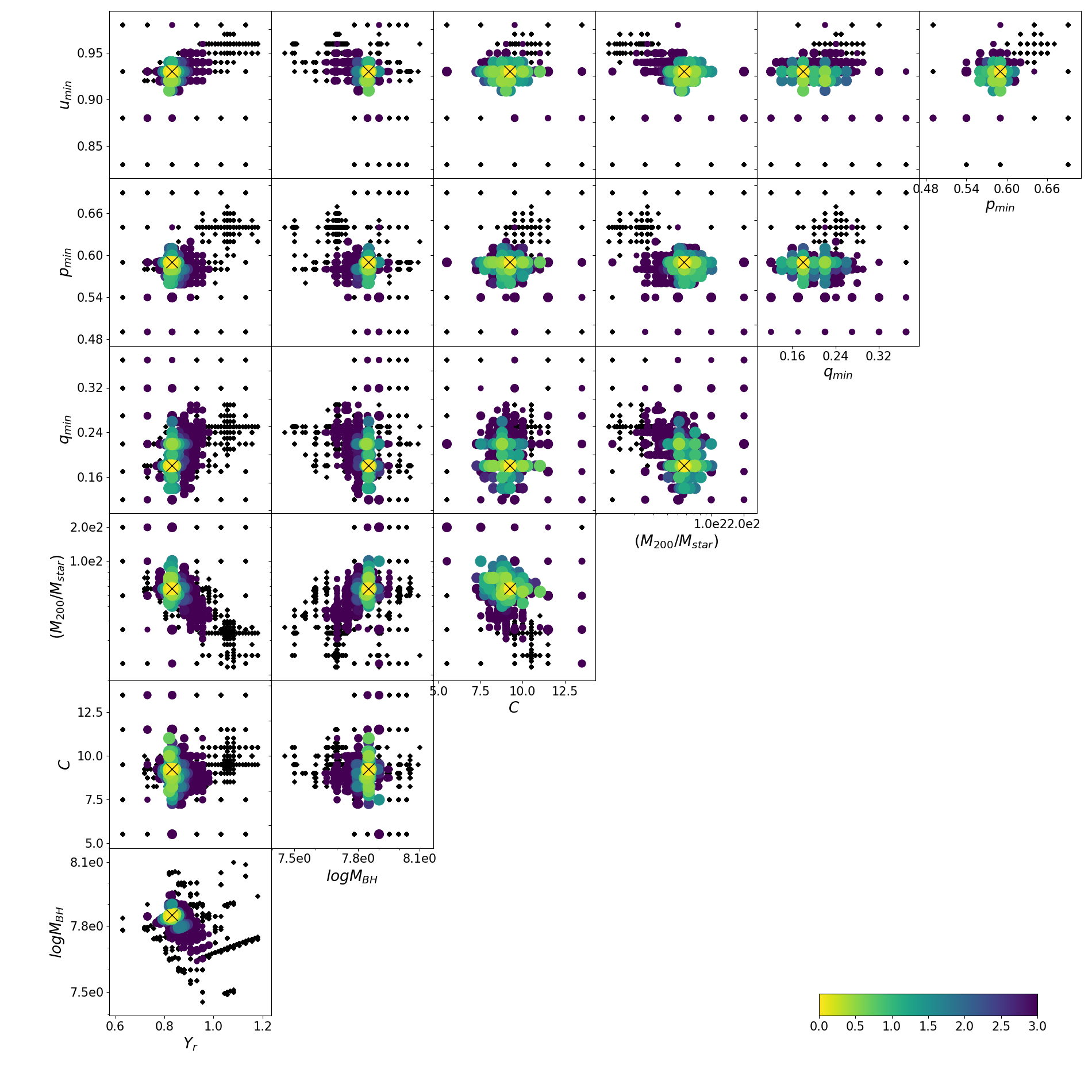}

      \caption{$\chi^2$ distribution of our dynamical models to constrain intrinsic shape and dark matter parameters for the mass model with varying $M_*/L$ and MW-like IMF. The seven hyper-parameters are the mass scaling factor S (dimensionless), the black hole mass in solar units, the dark matter concentration c, the dark matter halo mass in units of the stellar mass $M_{200}/M_{star}$, the intrinsic minor-to-major axis ratio q, the intrinsic medium-to-major axis ratio p, and the ratio between projected and intrinsic major axis u. Each point is one model that is colour-coded according to its $\chi^2$ value as shown in the colour bar, where ($\chi^2$-$\chi_{\rm min}^2)/\sqrt{2n_{\rm GH}\times N_{\rm kin}} < 1$ indicates the models within 1$\sigma$ confidence interval. Here, $n_{\rm GH}$ corresponds to the number of fitted Gauss-Hermite moments (4) and $N_{\rm kin}$ is the number of kinematic bins. The black cross marks the best-fitting model.
      }
      \label{ff:chi2_global2}
\end{figure*}

\begin{figure*}
  \centering
      \includegraphics[width=1\textwidth]{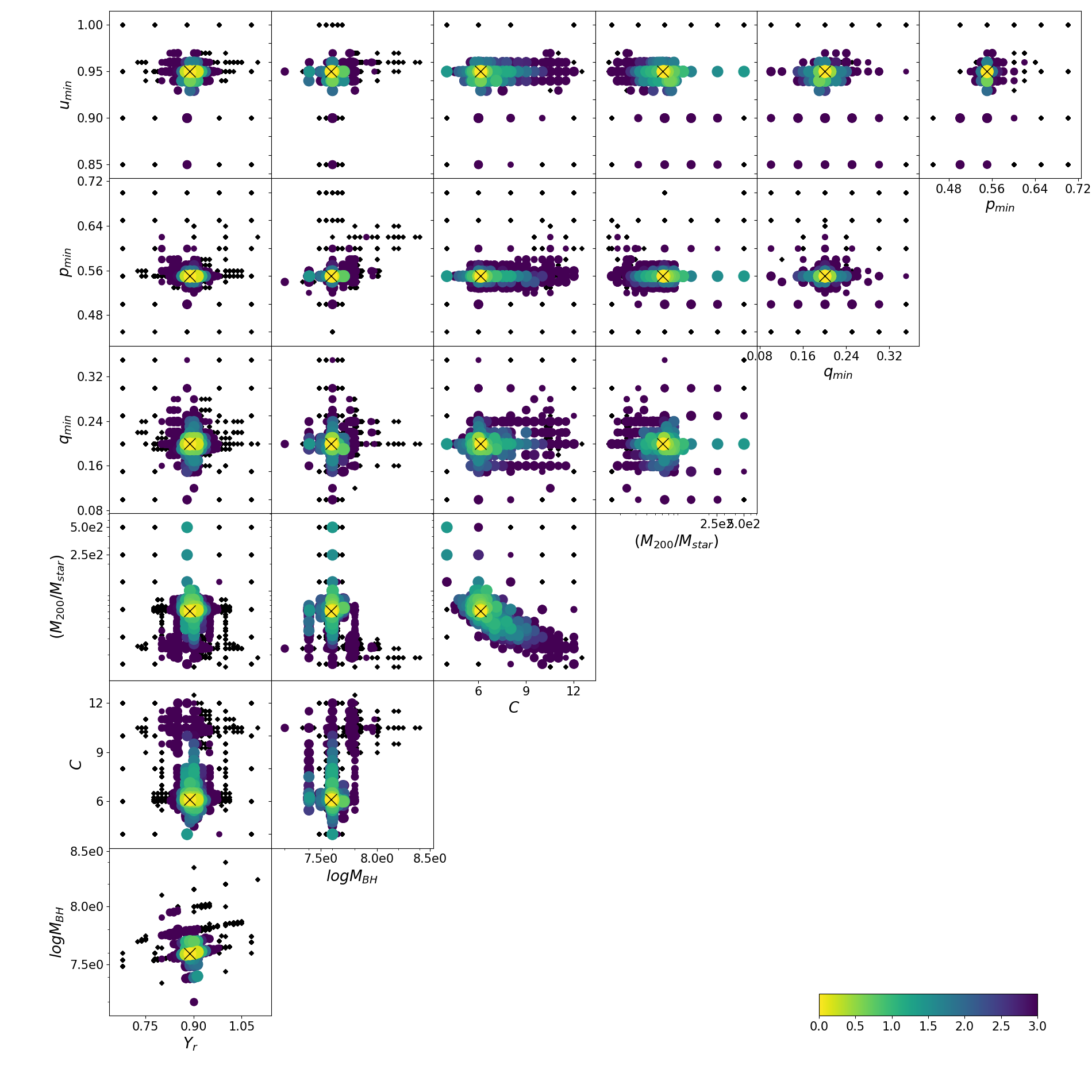}

      \caption{$\chi^2$ distribution of our dynamical models to constrain intrinsic shape and dark matter parameters for the mass model with varying $M_*/L$ and free IMF. The seven hyper-parameters are the mass scaling factor S (dimensionless), the black hole mass in solar units, the dark matter concentration c, the dark matter halo mass in units of the stellar mass $M_{200}/M_{star}$, the intrinsic minor-to-major axis ratio q, the intrinsic medium-to-major axis ratio p, and the ratio between projected and intrinsic major axis u. Each point is one model that is colour-coded according to its $\chi^2$ value as shown in the colour bar, where ($\chi^2$-$\chi_{\rm min}^2)/\sqrt{2n_{\rm GH}\times N_{\rm kin}} < 1$ indicates the models within 1$\sigma$ confidence interval. Here, $n_{\rm GH}$ corresponds to the number of fitted Gauss-Hermite moments (4) and $N_{\rm kin}$ is the number of kinematic bins. The black cross marks the best-fitting model.
      }
      \label{ff:chi2_global3}
\end{figure*}

\begin{figure*}
  \centering
      \includegraphics[width=0.69\textwidth]{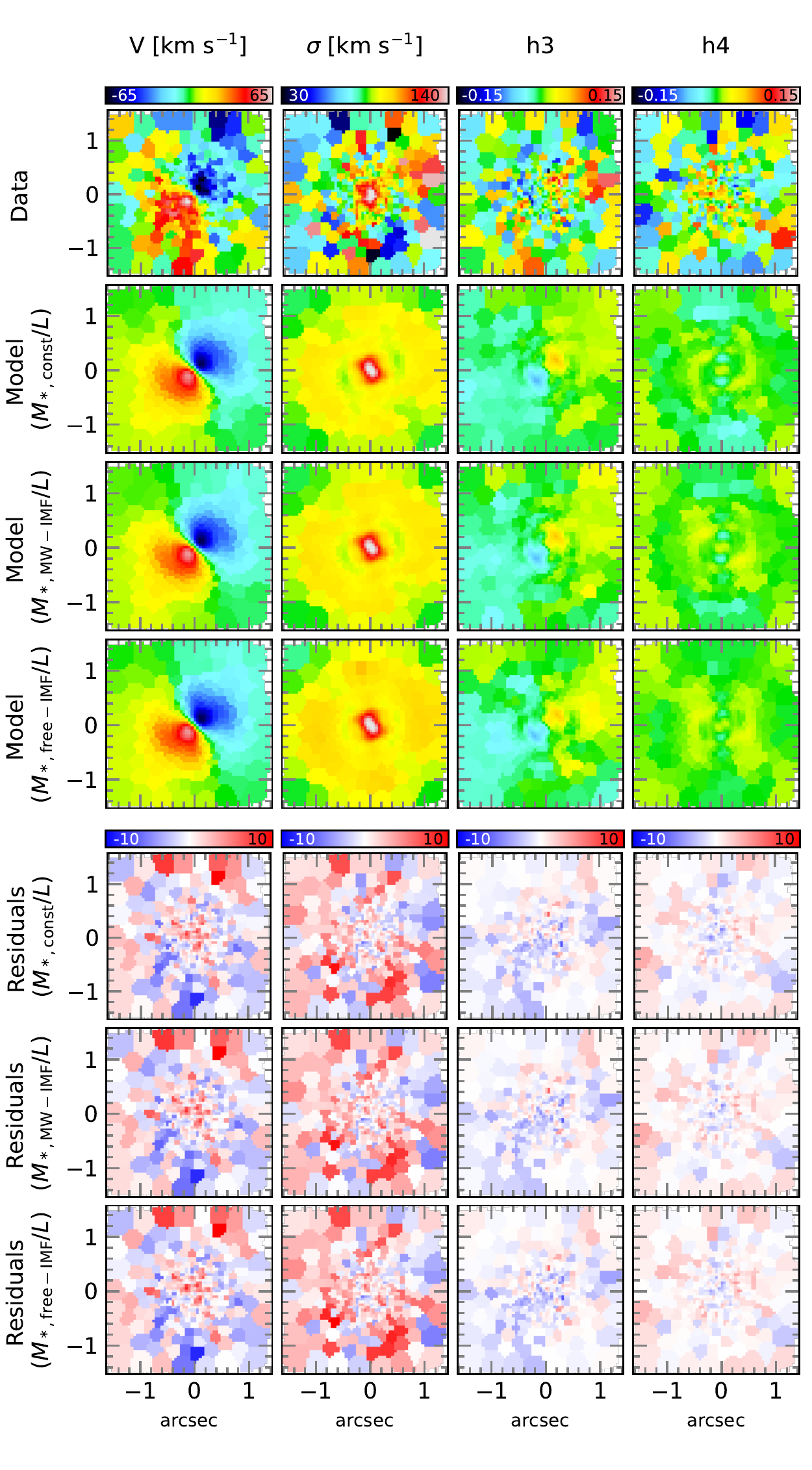}

      \caption{Comparison between SINFONI kinematics and best-fitting DYNAMITE models. From left to right: Mean velocity, velocity dispersion, $h_{3}$ and $h_{4}$ Gauss-Hermite moments. From top to bottom: Data,
best-fitting Schwarzschild model of the models with constant $M_*/L$, models with varying $M_*/L$ and constant IMF, and models with varying $M_*/L$ and varying IMF. The bottom three rows show the residual maps of the models with constant $M_*/L$, models with varying $M_*/L$ and constant IMF and models with varying $M_*/L$ and varying IMF. Residuals are defined as difference between the Schwarzschild model and observed kinematics divided by the observational errors. North is up and east to the left.
      }
      \label{ff:dyna_sinfoni}
\end{figure*}
\begin{figure*}
  \centering
      \includegraphics[width=0.69\textwidth]{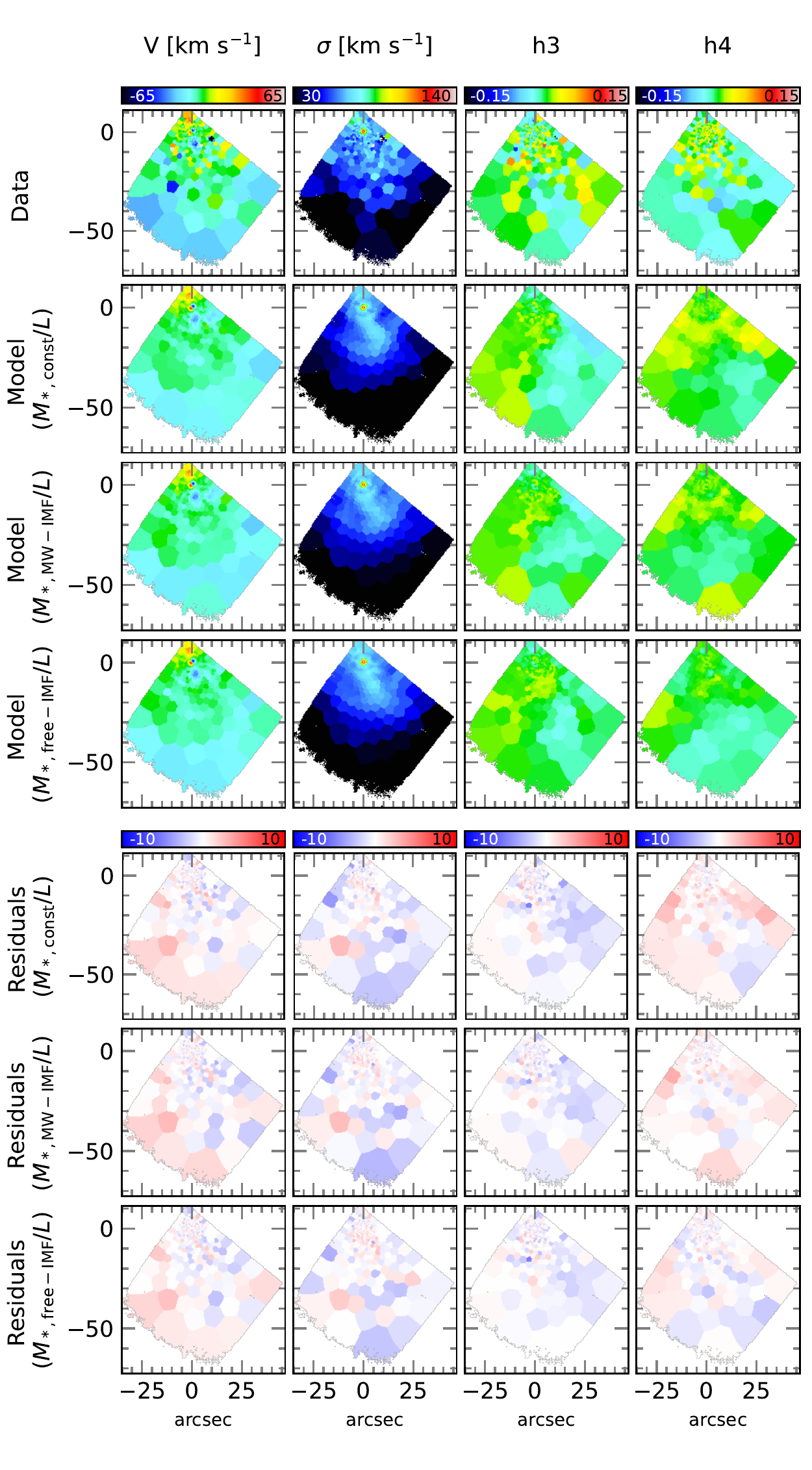}

      \caption{Comparison between MUSE kinematics and best-fitting DYNAMITE models. From left to right: Mean velocity, velocity dispersion, $h_{3}$ and $h_{4}$ Gauss-Hermite moments. From top to bottom: Data,
best-fitting Schwarzschild model of the models with constant $M_*/L$, models with varying $M_*/L$ and constant IMF, and models with varying $M_*/L$ and varying IMF. The bottom three rows show the residual maps of the models with constant $M_*/L$, models with varying $M_*/L$ and constant IMF and models with varying $M_*/L$ and varying IMF. Residuals are defined as difference between the Schwarzschild model and observed kinematics divided by the observational errors. North is up and east to the left.
      }
      \label{ff:dyna_muse}
\end{figure*}

\end{appendix}
\end{document}